\documentclass[%
 reprint,
 amsmath,amssymb,
 aps,
]{revtex4-2}

\usepackage{graphicx}
\usepackage{dcolumn}
\usepackage{bm}
\usepackage{hyperref} 
\usepackage{url}
\usepackage[utf8]{inputenc}
\usepackage{multirow}
\usepackage{braket}
\usepackage{amsmath}
\usepackage{amssymb}
\usepackage{amsthm}
\usepackage{latexsym}
\usepackage{epsfig}
\usepackage{xcolor}

\usepackage{braket}
\usepackage{xcolor}

\begin{document}

\preprint{APS/123-QED}

\title{Eberhard limit for photon-counting Bell tests and its utility in quantum key distribution}

\author{%
	Thomas~McDermott$^{1,2,*}$, Morteza~Moradi$^1$, Antoni~Mikos-Nuszkiewicz$^1$,\\ \& Magdalena~Stobi\'nska$^{1,2}$\\
	\textit{$^1$ Institute of Informatics, Faculty of Mathematics, Informatics and Mechanics, University of Warsaw, Banacha 2, 02-097 Warsaw, Poland}\\
	\textit{$^2$ Institute of Theoretical Physics, Faculty of Physics, University of Warsaw, Pasteura 5, 02-093 Warsaw, Poland}\\
	\textit{*tommcdee78@gmail.com}
}

\date{\today}

\begin{abstract}
Loophole-free Bell tests are essential if one wishes to perform device-independent quantum key distribution (QKD), since any loophole could be used by a potential adversary to undermine the security of the protocol. Crucial work by Eberhard demonstrated that weakly entangled two-qubit states have a far greater resistance to the detection loophole than maximally entangled states, allowing one to close the loophole with detection efficiency greater than 2/3. Here we demonstrate that this same limit holds for photon-counting CHSH Bell tests which can demonstrate non-locality for higher dimensional multiphoton states such as two-mode squeezed vacuum and generalized Holland-Burnett states. In fact, we show evidence that these tests are in some sense universal, allowing feasible detection loophole-free tests for any multiphoton bipartite state, as long as the two modes are well correlated in photon number. Additionally, by going beyond the typical two-input two-output Bell scenario, we show that there are also photon-counting CGLMP inequalities which can also match the Eberhard limit, paving the way for more exotic loophole-free Bell tests. Finally we show that by exploiting this increased loss tolerance of non maximally entangled states, one can increase the key rates and loss tolerances of QKD protocols based on photon-counting tests.\end{abstract}

\maketitle


\section{Introduction}

Loophole-free Bell tests are the gold standard of entanglement testing \cite{loopholefreespins, loopholefreephotons, significant, strongtest, brunner}. They provide definitive proof that multiple parties share quantum entanglement which can then be used for applications such as device-independent quantum key distribution (DI-QKD)\cite{acin2006, acin2007, pironio, acin2011, pironio2021, vidick, diqkd_review1, diqkd_review2}, quantum metrology\cite{metrology, metrology2, metrology3, monikametrology} and device-independent randomness generation\cite{pironioRNG, RNG, RNG2}. If instead there exists some loopholes, the proof is not definitive but relies on some extra assumptions. For example, the detection loophole is opened by filtering out some measurement results (usually when a photon is lost) i.e. performing post-selection. The proof of entanglement is then not definitive but based on additional fair-sampling assumptions \cite{fairsampling1, fairsampling2}. In particular, QKD without loophole-free Bell tests are then vulnerable to attacks by exploiting these extra security assumptions, and commercial devices have been routinely hacked as a result \cite{detectorblinding1, detectorblinding2}.

Closing the detection loophole is experimentally demanding and requires minimizing losses to increase the detection efficiency above some threshold. The crucial first steps towards experimental tests which closed the detection loophole were enabled by pivotal works by Eberhard, who demonstrated that using non-maximally entangled states can greatly lower the strict efficiency requirements \cite{eberhard}. Specifically, using a maximally entangled polarization Bell pair $\frac{1}{\sqrt{2}}(\ket{\textrm{HV}} + \ket{\textrm{VH}})$, the detection loophole is closed only for detector efficiencies $\eta > 2(\sqrt{2}-1) \approx 82.8\%$, whereas using a non-maximally entangled state $\sqrt{1-g^2} \ket{\textrm{HV}} + g \ket{\textrm{VH}}$, one can approach the limit $\eta > 2/3$ as $g \to 0$. Paradoxically, this is when the state approaches the separable $\ket{HV}$, so that it becomes less tolerant to other types of noise. In practice, the optimal value of $g$ then depends on the relative importance of different sources of imperfection. The first experiments exploiting the Eberhard inequality chose a value of $g = 0.297$ \cite{closedetectionloophole}.

Nowadays loophole-free Bell tests are becoming more common \cite{loopholefreespins, loopholefreephotons, significant, strongtest}, with the first fully DI-QKD demonstrations being performed \cite{diqkd}. However, until now all these Bell tests have been performed for 2-qubit polarization/spin singlet states and using the typical CHSH test based on polarization/spin measurements\cite{CHSH}, precluding the possibility of testing the entanglement of some more exotic states, such as multiphoton/qudit states. Bell tests for multiphoton states of light based on photon-counting experiments were first theorised in the 1990s\cite{bw0, bw1, bw2, bw3} and have since been expanded on in a number of papers \cite{bwtmsv, bwcglmp, detectingentanglement, brask}, including a proposal for DI-QKD based on these tests \cite{mycroftProposalDistributionMultiphoton2022}. These tests work for a variety of important states such as two-mode squeezed vacuum\cite{bwtmsv}, generalized Holland-Burnett states\cite{mycroftProposalDistributionMultiphoton2022} and hybrid-entangled Schr{\"o}dinger cat states\cite{ketterer}.

In this work we investigate whether these photon-counting tests form a universal test for entanglement in the photon-number degree of freedom. We demonstrate that any entangled state with high photon-number correlations can violate one of a number of Bell inequalities without employing any post-selection or filtering of results, thus having the possibility of closing the detection loophole. This requires low losses and high efficiency detectors, however we demonstrate that states which asymptotically approach the vacuum state have the maximum possible efficiency tolerance $\eta > 2/3$, matching that obtained by Eberhard for polarization/spin measurements. Moreover, this tolerance is not limited to the CHSH tests introduced in the original papers, but is also a property of the multiple output CGLMP inequalities\cite{cglmp}, opening up loophole-free Bell tests to both more exotic states and more exotic inequalities. Finally, we demonstrate that the Eberhard inequality found in this paper has utility for DI-QKD protocols, allowing one to increase both the loss tolerance and key rates using less than maximally entangled states.

\section{Photon-counting Bell tests}

\begin{figure}
    \centering
    \includegraphics[width=.7\linewidth]{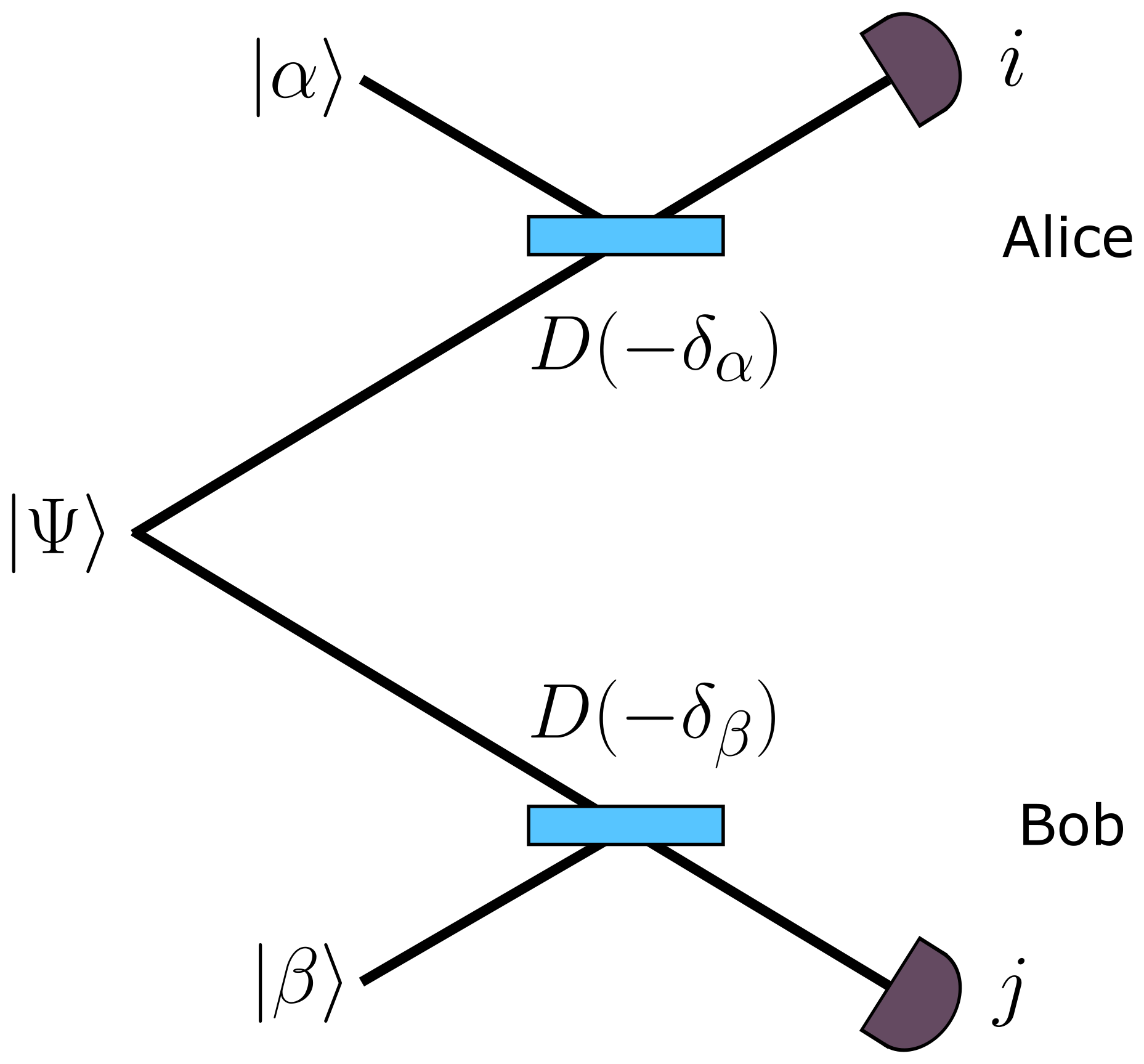}
    \caption{Setup for the photon-counting Bell tests. Alice and Bob share a two-mode state $\ket{\Psi}$ and wish to verify its entanglement. Each party interferes their mode with coherent states $\ket{\alpha}$ and $\ket{\beta}$, on beam splitters with reflectivities $r_a$, $r_b$ respectively, and then measures the transmitted photon-numbers with photon-number-resolving (PNR) detectors. By splitting their measurements into binary outcomes such as zero/non-zero photons or even/odd numbers of photons, they can evaluate the CHSH inequality for this scenario. Ideally the beam splitters are in the `displacement limit' with $r_a, r_b \to 0$ and $|\alpha|^2, |\beta|^2 \to \infty$ such that they perform the coherent displacements $D(-\delta_\alpha)$ and $D(-\delta_\beta)$, where $\delta_\alpha = i \alpha \sqrt{r}$, $\delta_\beta = i \beta \sqrt{r}$.}
    \label{fig1}
\end{figure}

Figure \ref{fig1} depicts a Bell test for a two-mode state of light $\ket{\Psi}$. The two modes are spatially separated, with one mode being sent to Alice and the other to Bob. Each party interferes their mode with a coherent state $\ket{\alpha}$ and $\ket{\beta}$, on variable beam splitters with reflectivities $r_a$ and $r_b$ respectively. Finally they use photon-number-resolving (PNR) detectors to measure the number of photons $i$, $j$ in the two transmitted modes behind the beam splitters. The combination of the complex amplitude $\alpha$ and reflectivity $r_a$ form a `measurement setting' for Alice, while $\beta$ and $r_b$ form the corresponding setting for Bob. In the limits $|\alpha|, |\beta| \to \infty$ and $r \to 0$ the beam splitter interactions are equivalent to displacement operators $D(-\delta_\alpha)$ and $D(-\delta_\beta)$ where $\delta_\alpha = i \alpha \arcsin(\sqrt{r_a}) \approx i \alpha \sqrt{r_a}$ and $\delta_\beta$ defined similarly. By collecting measurement statistics for $m \geq 2$ settings per party, Alice and Bobs can prove entanglement of their shared state, either by showing violation of a Bell inequality or via linear programming methods.

Since the coherent states have a finite probability to contain any number of photons, $i, j$ can be any integer, and the number of measurement outcomes is unbounded. To make use of known Bell inequalities it is thus necessary to reduce these to a finite number of outcomes, $\Delta$. The simplest case $m = 2, \Delta = 2$ has been analyzed in the previous literature \cite{bw0, bw1, bw2, bw3}. 
One way to reduce the output to two outcomes is for Alice (Bob) to assign a measurement outcome of $a= 1$ $(b=1)$ when zero photons are transmitted and an outcome of $a=-1$ $(b=-1)$ when non-zero photons are transmitted. The probability for a joint measurement of zero photons is then
\begin{equation}
	p(i=0,j=0) = \bra{\delta_\alpha}_a\bra{\delta_\beta}_b \ket{\Psi}\bra{\Psi} \ket{\delta_\alpha}_a\ket{\delta_\beta}_b = Q(\delta_\alpha, \delta_\beta),
\end{equation}
which is simply the two-mode Q-function of the state $Q(\delta_\alpha, \delta_\beta)$ (we neglect the usual factors of $\pi$). Moreover, the marginal probability that Alice detects zero photons 
\begin{equation}
	p(i = 0) = \bra{\delta_\alpha} \textrm{Tr}_b(\ket{\Psi}\bra{\Psi})\ket{\delta_\alpha} = Q_a(\delta_\alpha),
\end{equation}
is just the Q-function of her reduced state, $Q_a(\delta_\alpha)$, and Bob's probability $p(j = 0)$ is defined similarly. Here $\textrm{Tr}_b$ is a trace over Bob's modes. If Alice and Bob choose $\delta_\alpha$ and $\delta_\beta$ randomly from the sets $[\delta_{\alpha_1}, \delta_{\alpha_2}]$ and $[\delta_{\beta_1}, \delta_{\beta_2}]$ respectively, the CHSH inequality\cite{CHSH} for this zero/non-zero test can then be expressed entirely in terms of Q-functions as follows

\begin{equation}
    \frac{B_Q-2}{4} = Q_{11} + Q_{12} + Q_{21} - Q_{22} - Q_{a1} - Q_{b1},\label{bellQ}
\end{equation}
$|B_Q| \leq 2$, where $Q_{xy} = Q(\delta_{\alpha x}, \delta_{\beta y})$, $Q_{ax} = Q_a(\delta_{\alpha x})$ and $Q_{by} = Q_b(\delta_{\beta y})$.

Another simple way to obtain two outcomes is for Alice (Bob) to assign a measurement outcome of $a= 1$ $(b=1)$ when an even number of photons are transmitted and an outcome of $a=-1$ $(b=-1)$ when an odd number are transmitted. This even/odd test can also be directly related to a quasi-probability distribution, since the quantum correlation $\langle a b \rangle$ is given directly by the two-mode Wigner function
\begin{equation}
    W(\alpha, \beta) = \bra{\Psi} D(\alpha) D(\beta) (-1)^{a^\dagger a} (-1)^{b^\dagger b}  D(\alpha)^\dagger D(\beta)^\dagger \ket{\Psi}.
\label{Wdef}
\end{equation}
The CHSH inequality for this test is then
\begin{align}
	B_W = W_{11} + W_{12} + W_{21} - W_{22},\label{bellW}
\end{align}
$|B_W| \leq 2$, where $W_{xy} = W(\delta_{\alpha x}, \delta_{\beta y})$.

Perhaps the most simple way to generalize the above tests is to extend the number of outcomes $\Delta$ beyond two. This requires the following general formula to measure arbitrary photon numbers $i, j$
\begin{align}
p(i, j | \delta_{\alpha} \delta_{\beta}) &= \left|\bra{i}\bra{j} D(\delta_\alpha)D(\delta_\beta) \ket{\Psi}\right|^2 \label{pil}\\
&= \left|\sum_{m,n=0}^{\infty} c_{mn} D_{im}(\delta_\alpha) D_{jn}(\delta_\beta)\right|^2,\nonumber
\end{align}
where $D_{im}(\delta_\alpha) = \bra{i}D(\delta_\alpha)\ket{m}$ is the transition element of the displacement operator between two Fock states
\begin{align}
D_{im}&(\delta_\alpha) = \sqrt{m! i!} \hspace{0.4em} \delta_\alpha^{i-m} \hspace{0.4em} e^{-\frac{|\delta_\alpha|^2}{2}} \\
&\times \sum_{p=\text{max}(0,m-i)}^m \frac{(-|\alpha|^2)^p}{p! (m-p)! (p-m+i)!}.\nonumber
\end{align}

Using Eq. (\ref{pil}) we can evaluate inequalities in the scenario $m=2$, $\Delta = 3$. The facet-defining inequalities here are the CGLMP inequalities\cite{cglmp}, plus the usual projections onto CHSH. Instead of separating into zero/non-zero outcomes as before, we can now separate into zero photons, single photons and two or more photons, forming our three outcomes per party. The CGLMP inequality for this scenario may be written as
\begin{equation}
	I = G(\delta_{\alpha 1}, \delta_{\beta 1}) - G(\delta_{\alpha 2}, \delta_{\beta 1}) + G(\delta_{\alpha 2}, \delta_{\beta 2}) + G(\delta_{\beta 2}, \delta_{\alpha 1}),
\end{equation}
$I \leq 2$, where
\begin{align}
	G(\delta_{\alpha}, \delta_{\beta}) = &3\Big[p(i=0,j=0|\delta_\alpha,\delta_\beta) \\&+ p(i=1,j=0|\delta_\alpha,\delta_\beta)\nonumber\\ &+ p(i=1,j=1|\delta_\alpha,\delta_\beta)\Big]\nonumber\\ &-2p(i=1|\delta_\alpha) -2p(j=0|\delta_\beta)\nonumber\\&- p(i=0|\delta_{\alpha}) - p(j = 1|\delta_{\beta}) + 1.\nonumber
\end{align}

Notice that all of the described photon-counting Bell tests take into account every possible experimental outcome, including cases where photons are lost, so that no post-selection or fair sampling assumptions are required. This allows the possibility for these tests to close the detection loophole without any modifications.

\begin{figure*}
    \centering
    \includegraphics[width=\linewidth]{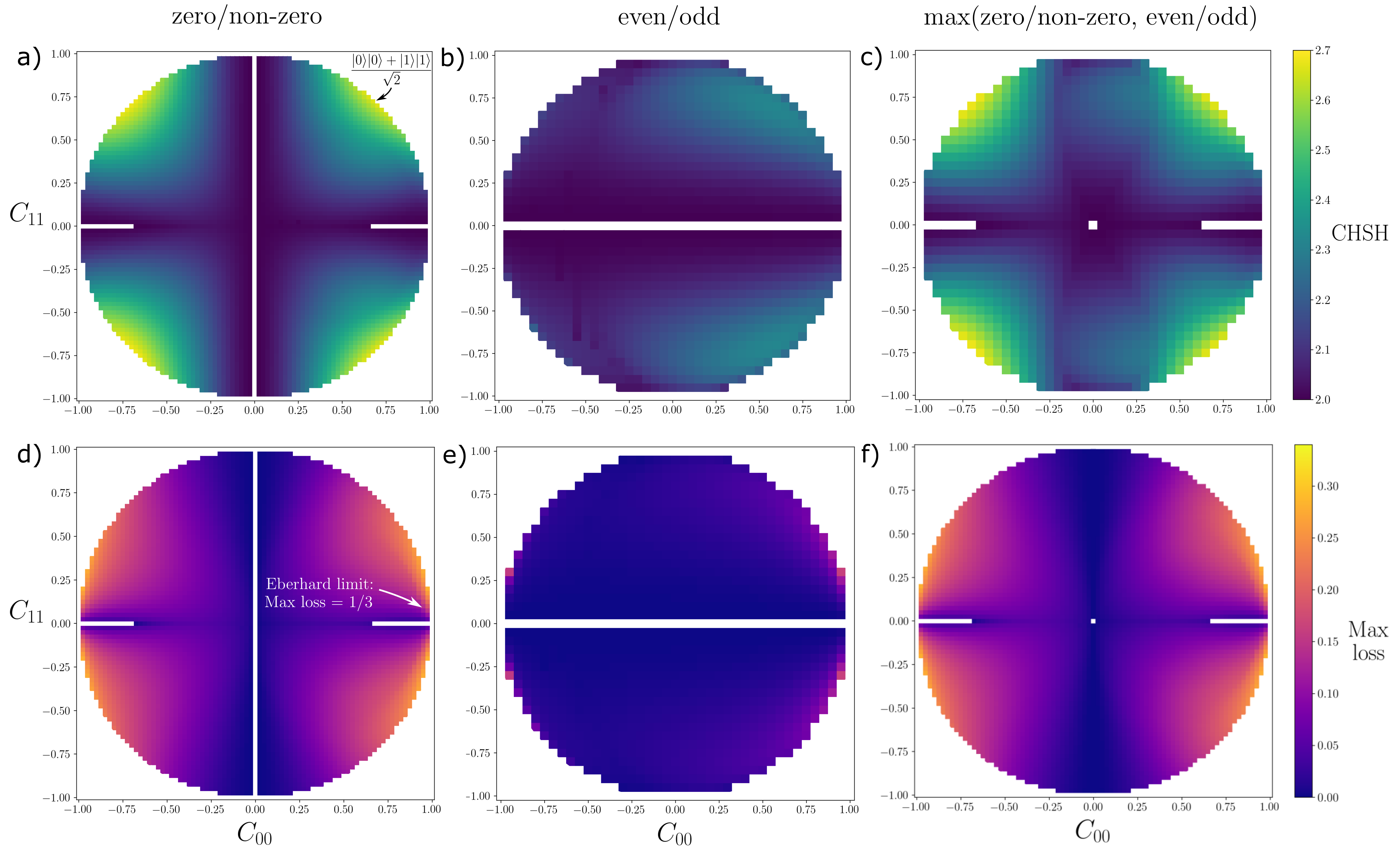}
    \caption{Photon-counting CHSH tests for highly correlated states of the form $C_{00}\ket{00} + C_{11}\ket{11} + C_{22}\ket{22}$, where $C_{22} = +\sqrt{1-C_{00}^2-C_{11}^2}$ is fixed by normalization. Figures (a), (b) and (c) show for each value $C_{00}, C_{11}$ the maximum violation for the zero/non-zero test, the even/odd test, and their combination, when numerically optimized over all measurement settings. The white areas show regions where there was no violation i.e. $\textrm{CHSH} \leq 2$. The greatest violation of 2.69 is found for the states $(\pm \ket{00} \pm \ket{11})/\sqrt{2}$. Figures (d), (e) and (f) shows the maximum loss before the CHSH test fails ($B < 2$), for the tests in (a), (b) and (c) respectively. The zero/non-zero test is significantly more loss tolerant than the even/odd test, particularly in the vicinity of the points $|C_{00}| \approx 1$ where the state approaches vacuum. Here there is an Eberhard level of loss tolerance $33\%$, although the loss tolerance is significant even outside of this extreme.}
    \label{fig2}
\end{figure*}

\begin{figure*}
    \centering
    \includegraphics[width=\linewidth]{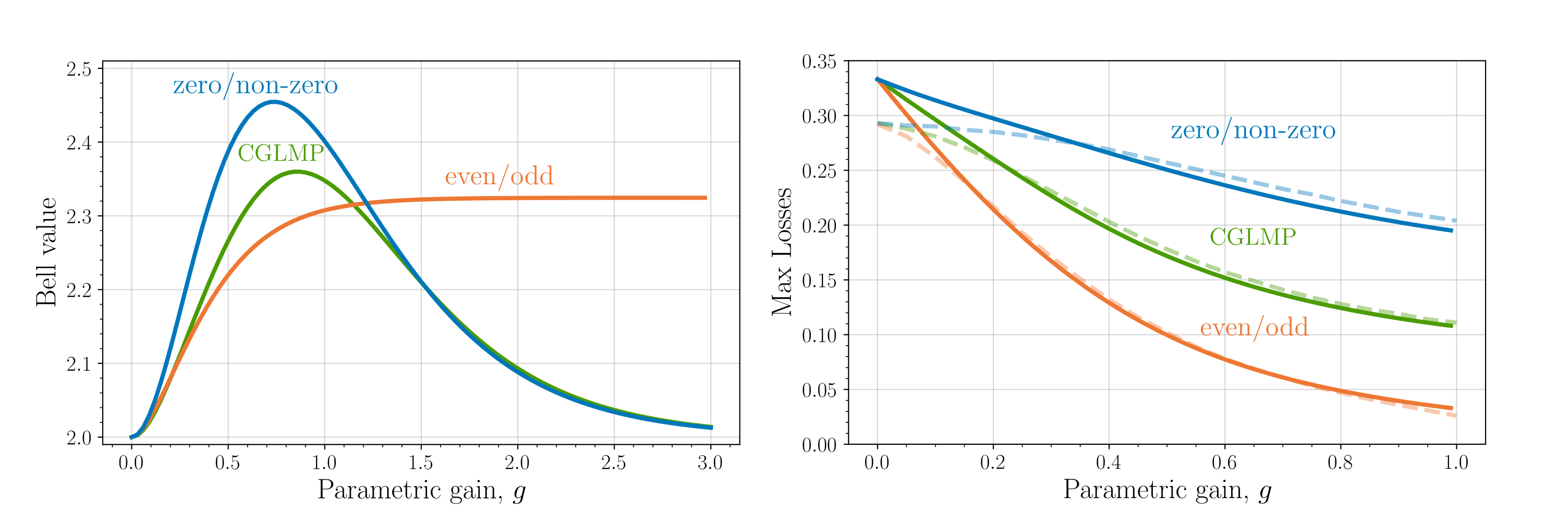}
    \caption{Photon-counting Bell tests on a two-mode squeezed vacuuum state as a function of parametric gain. (a) Bell inequality value for three different tests. (b) Maximum loss tolerance for these tests, showing Eberhard level loss tolerance $\eta > 2/3$ as the parametric gain tends to zero. All three tests satisfy this requirement although the even/odd and CGLMP tests degrade much faster with increasing gain.}
    \label{fig3}
\end{figure*}

To determine whether such tests work as a universal test for entanglement-verification in the photon number regime, we consider states of the form $\sum_{m,n=0}^d C_{mn} \ket{m}\ket{n}$. We first considered the perfectly correlated states $C_{mn} = C_{nn} \delta_{mn}$, and performed an exhaustive search over all complex values $C_{nn}$ for $d \leq 4$. By numerically maximizing the CHSH violation over all four complex displacements, we found that the test worked to prove entanglement for the vast majority of states, the only exception being states with very weak levels of entanglement i.e. those that are close to separable states. Thus we may conjecture that the photon-counting Bell tests work for all perfectly correlated pure entangled states, and since these tests do not require any post-selection or filtering of results, they have the possibility of closing the detection loophole. To illustrate this point, we plot in Fig. 2 the results for states of up to $d = 2$ photons, with real coefficients $C_{nn}$. $C_{00}$ and $C_{11}$ are varied while $C_{22}$ is fixed by the normalization condition $C_{22} = +\sqrt{1-C^2_{00} -C^2_{11}}$. The positive square root can be taken without loss of generality since the negative square root is equivalent to flipping the sign of all other coefficients, as the state would change only by a global phase. 

Fig. \ref{fig2}(a) shows the CHSH value for the zero/non-zero test. One observes that the test succeeds mostly everywhere, notably except along the line $C_{00} = 0$, i.e. states which have no vacuum component.  The test also fails for states like $C_{00} + \epsilon C_{22}$, $\epsilon \ll 1$ i.e. states that are close to vacuum with no single photon component. Everywhere else this test works well, and can yield fairly high violations of local realism. The greatest violation of 2.69 is found for the states $(\pm \ket{00} \pm \ket{11})/\sqrt{2}$. Although these states are maximally entangled in 2 dimensions, the basis states are not qubits but Fock states which are infinite dimensional. Fig. \ref{fig2}(b) shows the CHSH value for the even/odd test, where one observes that it is much weaker than the zero/non-zero test. One advantage however is that, unlike the zero/non-zero test, it can succeed for states with no vacuum component $C_{00} = 0$. This test instead fails for states with a definite photon number parity e.g. along the line $C_{11} = 0$, as could be expected. Performing both of these tests can then eliminate the weaknesses of the individual tests, leading to fairly large CHSH violations for the vast majority of photon-number correlated states, as shown in Fig. \ref{fig2}(c). In contrast, states without perfect correlation, $C_{mn} \neq 0$ for $m \neq n$, cannot always be verified by these tests. In general, as the levels of entanglement and photon-number correlations increase, the more likely the test will succeed. 

\section{Loss Tolerance}

The amount of CHSH violation does not give a complete picture, however. A crucial factor for experimental tests is how tolerant the Bell inequality violation is to photonic losses. These losses may be modelled by interfering the state with vacuum on a fictional beamsplitter of reflectivity $1-\eta$ and tracing out the reflected beam, equivalent to a photon being successfully transmitted with probability $\eta$. With these losses considered solely to occur after the displacement operation, $\eta$ can be interpreted as the detection efficiency of the two detectors. Fig. \ref{fig2}(d) shows the loss tolerance of the zero/non-zero test for the 2-photon states. Interestingly, one observes that as one approaches the vacuum state $C_{00} \to 1$, the loss tolerance approaches $\eta > 2/3$, exactly the Eberhard limit. Placing the losses in the source i.e. before the displacement operation, leads to very similar conclusions with only a slightly more strict limit $\eta > 0.71$. From a mathematical point of view the fact that we can reach the Eberhard limit is not too surprising, since the state in this limit can be approximated as something like $\sqrt{1-g^2} \ket{00} + g\ket{11}$ which is mathematically identical to the 2-qubit state originally considered by Eberhard. However, it is interesting that this limit can be approached through this completely different measurement scheme, with routine optical experimental devices\cite{wignerbypnr, wignerbypnr2}, and with entanglement in photon-number rather than polarization or spin. The even/odd test has a much worse loss tolerance, as seen in Fig. \ref{fig2}(e). Although it also displays Eberhard level loss tolerance in the same limit, it drops rapidly as we leave the vicinity of the vacuum state $\ket{00}$, so that it is not really visible in this figure.

For a concrete example we can consider the two-mode squeezed vacuum state $\ket{\psi} = \frac{1}{\cosh g}\sum_{n=0}^\infty (-e^{i\phi} \tanh g)^n \ket{n}\ket{n}$ which is a qudit state of in principle infinite dimension. The photon-counting Bell tests have been considered for these states in the early papers\cite{bw0, bw1, bw2, bw3}, althought they didn't find the greatest violation of the Bell inequalities and didn't realise its large loss tolerance in the low squeezing limit $g \to 0$.

The relevant Q-functions for the TMSV state are
\begin{equation}
    Q(\alpha, \beta) = \frac{1}{\cosh^2 g}\exp[-|\alpha|^2-|\beta|^2-2\textrm{Re}\{\alpha \beta e^{-i\phi}\} \tanh g],
\end{equation}
\begin{equation}
    Q_a(\alpha) = \frac{1}{\cosh^2 g} \exp[-|\alpha|^2(1-\tanh^2 g)],
\end{equation}
with $Q_b(\beta)$ defined analogously, while the two-mode Wigner function is
\begin{align}
    W(\alpha, \beta) = \exp[&-2(|\alpha|^2+|\beta|^2)\cosh(2g)\\&-4\textrm{Re}\{ \alpha\beta e^{-i\phi}\} \sinh(2g)].
\end{align}
The CHSH parameters for both tests, $|B_Q|$ and $|B_W|$ were numerically maximized by a combination of differential evolution, Nelder-mead and simulated annealing methods, using Eqs. (\ref{bellQ}) and (\ref{bellW}). For the zero/non-zero test, the optimal choice of measurement settings $[\delta_{\alpha_1}, \delta_{\alpha_2}], [\delta_{\beta_1}, \delta_{\beta_2}]$ maximizing $|B_Q|$ were found to have the form $[x, -y], [x e^{i\phi}, -ye^{i\phi}]$, where $x$ and $y$ are positive real numbers which depend on the amount of squeezing $g$. This choice of relative phases can be understood as those maximizing $Q(\delta_{\alpha 1}, \delta_{\beta 2})$ and $Q(\delta_{\alpha 2}, \delta_{\beta 1})$ while minimizing $Q(\delta_{\alpha 2}, \delta_{\beta 2})$. The relative phase between $\delta_{\alpha_1}$ and $\delta_{\alpha_2}$ is then fixed, forcing $Q(\delta_{\alpha 1}, \delta_{\beta 1})$ to be minimized. For the even/odd test, the optimal settings maximizing $|B_W|$ have the same form, with similar reasoning, in this case maximizing/minimizing the various Wigner functions rather than the Q functions.

The results are shown in Fig. \ref{fig3}(a) as a function of parametric gain, $g$. One sees that the zero/non-zero test is optimal for the more experimentally relevant low gain regime, shown by the fact that $|B_Q| > |B_W|$, reaching an optimal value of $|B_Q| = 2.45$ at $g = 0.74$. Increasing $g$ further, the zero/non-zero test weakens due to the fact that zero photon events become increasing unlikely. The even/odd test dominates here $|B_W| > |B_Q|$, and reaches a maximum value $|B_W| = 2.32$ in the limit of high gain $g \to \infty$. Since the TMSV state becomes the original EPR state in this limit, in theory the even/odd test becomes a solution to the EPR paradox as originally proposed. However, as pointed out in the original analysis \cite{bw1}, there is difficulty both reaching, and performing the relevant measurements in, this limit. These results further highlight the general features we noticed before. The zero/non-zero test works better for low photon-number states i.e. low squeezing $g$, and the even/odd test works better for high photon-number states i.e. high squeezing $g$. Additionally, we also consider the CGLMP inequality with 3 outputs, distinguishing between 0, 1 and 2 or more photons. This performs similarly to the zero/non-zero test, although the Bell inequality violation is slightly lower. Furthermore from Fig. \ref{fig3}(b), one sees that all three tests reach the Eberhard level loss tolerance $\eta > 2/3$, although the zero/non-zero test has a loss tolerance which decays much less rapidly than the other two, so is more practical for testing the non-locality of the two-mode squeezed vacuum state. The reason for this is that losing a photon is catastrophic for the even/odd test since it changes an even result to an odd one and vice-versa. In contrast the zero/non-zero test is more robust since a non-zero result can remain non-zero. The fact that the CGLMP inequality also displays Eberhard level loss tolerace in the low squeezing regime, opens up the possibility of loophole-free Bell inequality tests using higher dimensional inequalities. Unfortunuately, the loss tolerance degrades faster with parametric gain than the zero/non-zero CHSH test, so that it appears to have limited practical use for the two-mode squeezed vacuum state in particular.

The Eberhard limit is reached when we consider detector losses i.e. with the losses occurring after the displacement operation, however we have also considered losses in the source i.e. with losses occurring before the displacement operation. The loss tolerance in this case is shown as the dashed lines in Fig. \ref{fig3}(b). The conclusions are not much changed, except that we reach a slightly worse maximum loss tolerance of $\eta > 0.71$.

\section{Utility for QKD}

As an example of the utility of this Eberhard limit for photon-counting, we can consider a DI-QKD protocol based upon these tests, similar to what has been considered in\cite{mycroftProposalDistributionMultiphoton2022}. There, a proposal was considered in which generalized Holland-Burnett states (i.e. the output of two Fock states interfering on a beam splitter) are distributed between two parties, and the zero/non-zero Bell test was used to establish a shared secret key. We now consider the non-maximally entangled state $\sqrt{1-\epsilon}\ket{00} + \sqrt{\epsilon} \ket{11}$ and demonstrate that using non-maximal entanglement can yield increased key rates and loss tolerances for key distribution. For $\epsilon = 1/2$ this state becomes $\frac{1}{\sqrt{2}}(\ket{00}+\ket{11})$ which leads to the same Bell inequality violation $~ 2.69$ and loss tolerance $~ 15\%$ as the state considered in \cite{mycroftProposalDistributionMultiphoton2022}. We have already seen that lowering $\epsilon$ from this symmetrical value can increase the loss tolerance of the Bell tests, approaching the Eberhard limit, but it remains to be seen whether it can also increase the loss tolerance of the DI-QKD protocol as a whole, since in general the loss tolerance of DI-QKD is even more strict than the loophole-free Bell test itself.

We consider a simple DI-QKD scenario based on CHSH violation as outlined in\cite{acin2006, acin2007, pironio, vidick}. In addition to the two measurement settings per party used in the CHSH test $[\delta_{\alpha_1}, \delta_{\alpha_2}]$ and $[\delta_{\beta_1}, \delta_{\beta_2}]$, Alice now uses a third setting $\delta_{\alpha_0}$. Each time the entangled state is distributed to them, Alice performs a measurement randomly choosing between her three settings, and Bob randomly chooses between his two settings. After a number of trials $N \to \infty$, they publicly communicate the settings used for each trial. A number of results using settings from the sets $[\delta_{\alpha_1}, \delta_{\alpha_2}]$ and $[\delta_{\beta_1}, \delta_{\beta_2}]$ are also communicated publicly and used to evaluate the CHSH inequality, while a small sample, $N'$ of results using the combination $\delta_{\alpha_0}$ and $\delta_{\beta_1}$ are kept secret and form the raw key shared between Alice and Bob. Violation of the CHSH inequality guarantees that Eve cannot obtain full information about the key, providing information-theoretic security of communications where they key is used in a one-time-pad.

After performing classical privacy amplification and error correction on their raw key of length $N'$, they obtain the final key of reduced length $K N'$ where the key rate $K$ is given by\cite{devetakwinter}
\begin{equation}
	K = I(A_0:B_1) - I(B_1:E).
\end{equation}
Here $I(A_0:B_1)$ is the mutual information between Alice and Bob (when using settings  $\delta_{\alpha_0}, \delta_{\beta_1}$), specifying the fraction of bits that are kept after performing error correction, while $I(B_1:E)$ is the mutual information between Bob and Eve, giving the number of bits that are sacrificed for privacy amplification. Assuming Eve performs optimal collective attacks against the protocol, the key rate can be lower bounded\cite{pironio, acin2011, pironio2021} by
\begin{equation}
	K = 1 - h\left(\frac{1+\sqrt{B^2/4-1}}{2}\right) - H(A_0|B_1),\label{keyrate}
\end{equation}
where $h(x)$ is the binary entropy function, $B$ is the obtained CHSH inequality value, and $H(A_0|B_1)$ is the conditional entropy between Alice and Bob. There are also methods for upgrading security against collective attacks to security against the most general coherent attacks \cite{vidick}. Having loophole-free Bell inequality violation $B > 2$ is required, but not sufficient, to obtain a finite key rate $K > 0$, since the entropy $H(A_0|B_1)$ must also be minimized. For this reason, we cannot obtain a loss tolerance of $\eta > 2/3$ for DI-QKD, but by using non-maximally entangled states, the loss tolerance can still be improved.

\begin{figure}
    \centering
    \includegraphics[width=\linewidth]{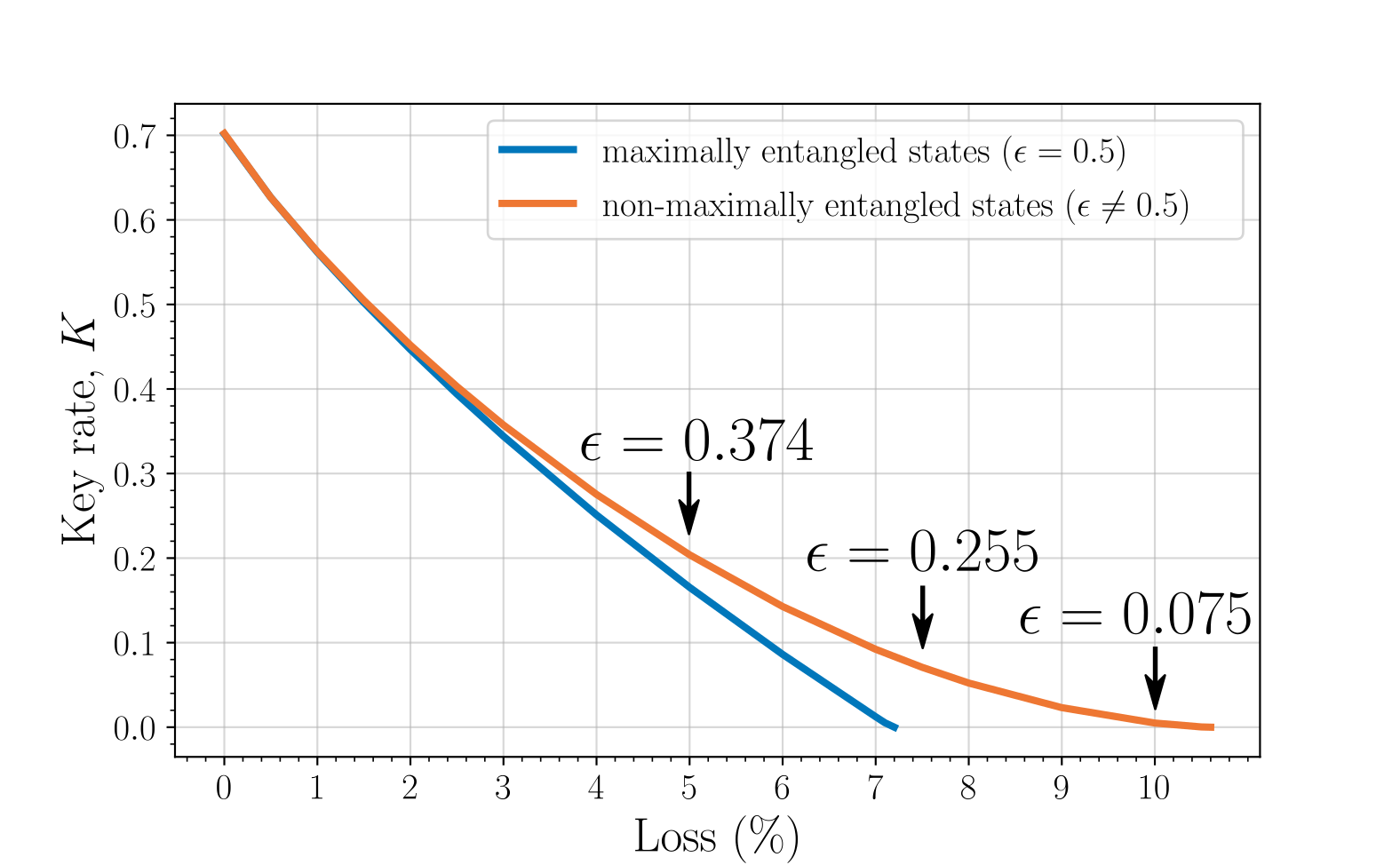}
    \caption{Demonstration of the utility of the increased loss tolerance of less maximally entangled states for photon-counting QKD demonstrations. Both the key rate and loss tolerance can be increased by using non-maximally entangled states $\sqrt{1-\epsilon}\ket{00} + \sqrt{\epsilon}\ket{11}$ rather than the maximally entangled states (in two-dimensions).} 
    \label{fig4}
\end{figure}

By numerically maximizing Eq. (\ref{keyrate}) with respect to the three settings of Alice and two settings of Bob, we have calculated the key rate that Alice and Bob may obtain by distributing the state $\sqrt{1-\epsilon}\ket{00} + \sqrt{\epsilon} \ket{11}$, for varying levels of photonic loss. The results for the maximally entangled state $\epsilon = 1/2$ are shown in the blue line Fig. \ref{fig4}, where one observes that after 7\% loss, the key rate drops to zero, indicating a rather strict efficiency requirement of 93\%. However, this can be improved considerably by instead distributing non-maximally entangled states. For each value of loss we have found the optimal value of $\epsilon$ and calculated the key rates, which are shown as the orange line in Fig. \ref{fig4}. Compared to maximally entangled states, the loss tolerance is increased from 7\% to 10.5\%, a 50\% improvement.

\section{Conclusion}

We have demonstrated the Eberhard inequality for photon-counting Bell tests, showing that detection loophole-free Bell tests are possible with efficiencies as low as 2/3, for photon-number entangled states which are generally multiphoton. We have shown this for two types of CHSH tests, based on discriminating between zero/non-zero photons, and even/odd numbers of photons respectively, but we have also shown it for CGLMP inequalities with more than two outputs. This opens up loophole-free Bell tests, and applications such as QKD, to more exotic states and types of entanglement than typical polarization/spin singlet states. The Eberhard limit for CGLMP inequalities could also allow detection loophole-free tests other than CHSH for the first time. The CHSH tests considered have been shown to work for states that are perfectly correlated in photon number, although for states without perfect correlation it can sometimes fail. Generally the more correlated in photon number states are, the more likely they are to violate these photon counting Bell tests, although a complete set of criteria to understand which states should violate these tests is still sought after. 

To highlight the utility of the Eberhard level of loss tolerance for photon-counting Bell tests, we have shown that the loss tolerances and key rates of DI-QKD protocols can be drastically improved by the distribution of photon-number entangled states which are not maximally entangled. DI-QKD represents the ultimate level of cryptography which is immune to any eavesdropping at either the source or detection, and DI-QKD proposals based on photon number entanglement is an active area of research\cite{mycroftProposalDistributionMultiphoton2022} since it has been shown to yield high key rates and immunity to large transmission losses. Thus our work represents a key step toward realization of this ambitious goal.

\begin{acknowledgments}
T.M. and M.S. were supported by the Foundation for Polish Science ``First Team'' project No. POIR.04.04.00-00- 220E/16-00 (originally FIRST TEAM/2016-2/17). A.M.-N. and M.S. were supported by the National Science Centre ``Sonata Bis'' project No. 2019/34/E/ST2/00273. M.M. and M.S. were supported by the European Union’s Horizon 2020 research and innovation programme under the Marie Skłodowska-Curie project ``AppQInfo'' No. 956071. T.M. and M.S. were supported by the National Science Centre within the QuantERA II Programme that has received funding from the European Union’s Horizon 2020 research and innovation programme under Grant Agreement No 101017733, project ``PhoMemtor'' No. 2021/03/Y/ST2/00177.
\end{acknowledgments}


\bibliography{eberhard}

\begin{thebibliography}{42}%
\makeatletter
\providecommand \@ifxundefined [1]{%
 \@ifx{#1\undefined}
}%
\providecommand \@ifnum [1]{%
 \ifnum #1\expandafter \@firstoftwo
 \else \expandafter \@secondoftwo
 \fi
}%
\providecommand \@ifx [1]{%
 \ifx #1\expandafter \@firstoftwo
 \else \expandafter \@secondoftwo
 \fi
}%
\providecommand \natexlab [1]{#1}%
\providecommand \enquote  [1]{``#1''}%
\providecommand \bibnamefont  [1]{#1}%
\providecommand \bibfnamefont [1]{#1}%
\providecommand \citenamefont [1]{#1}%
\providecommand \href@noop [0]{\@secondoftwo}%
\providecommand \href [0]{\begingroup \@sanitize@url \@href}%
\providecommand \@href[1]{\@@startlink{#1}\@@href}%
\providecommand \@@href[1]{\endgroup#1\@@endlink}%
\providecommand \@sanitize@url [0]{\catcode `\\12\catcode `\$12\catcode
  `\&12\catcode `\#12\catcode `\^12\catcode `\_12\catcode `\%12\relax}%
\providecommand \@@startlink[1]{}%
\providecommand \@@endlink[0]{}%
\providecommand \url  [0]{\begingroup\@sanitize@url \@url }%
\providecommand \@url [1]{\endgroup\@href {#1}{\urlprefix }}%
\providecommand \urlprefix  [0]{URL }%
\providecommand \Eprint [0]{\href }%
\providecommand \doibase [0]{https://doi.org/}%
\providecommand \selectlanguage [0]{\@gobble}%
\providecommand \bibinfo  [0]{\@secondoftwo}%
\providecommand \bibfield  [0]{\@secondoftwo}%
\providecommand \translation [1]{[#1]}%
\providecommand \BibitemOpen [0]{}%
\providecommand \bibitemStop [0]{}%
\providecommand \bibitemNoStop [0]{.\EOS\space}%
\providecommand \EOS [0]{\spacefactor3000\relax}%
\providecommand \BibitemShut  [1]{\csname bibitem#1\endcsname}%
\let\auto@bib@innerbib\@empty
\bibitem [{\citenamefont {Hensen}\ \emph {et~al.}(2015)\citenamefont {Hensen},
  \citenamefont {Bernien}, \citenamefont {Dr{\'e}au}, \citenamefont {Reiserer},
  \citenamefont {Kalb}, \citenamefont {Blok}, \citenamefont {Ruitenberg},
  \citenamefont {Vermeulen}, \citenamefont {Schouten}, \citenamefont
  {Abell{\'a}n}, \citenamefont {Amaya}, \citenamefont {Pruneri}, \citenamefont
  {Mitchell}, \citenamefont {Markham}, \citenamefont {Twitchen}, \citenamefont
  {Elkouss}, \citenamefont {Wehner}, \citenamefont {Taminiau},\ and\
  \citenamefont {Hanson}}]{loopholefreespins}%
  \BibitemOpen
  \bibfield  {author} {\bibinfo {author} {\bibfnamefont {B.}~\bibnamefont
  {Hensen}}, \bibinfo {author} {\bibfnamefont {H.}~\bibnamefont {Bernien}},
  \bibinfo {author} {\bibfnamefont {A.~E.}\ \bibnamefont {Dr{\'e}au}}, \bibinfo
  {author} {\bibfnamefont {A.}~\bibnamefont {Reiserer}}, \bibinfo {author}
  {\bibfnamefont {N.}~\bibnamefont {Kalb}}, \bibinfo {author} {\bibfnamefont
  {M.~S.}\ \bibnamefont {Blok}}, \bibinfo {author} {\bibfnamefont
  {J.}~\bibnamefont {Ruitenberg}}, \bibinfo {author} {\bibfnamefont {R.~F.~L.}\
  \bibnamefont {Vermeulen}}, \bibinfo {author} {\bibfnamefont {R.~N.}\
  \bibnamefont {Schouten}}, \bibinfo {author} {\bibfnamefont {C.}~\bibnamefont
  {Abell{\'a}n}}, \bibinfo {author} {\bibfnamefont {W.}~\bibnamefont {Amaya}},
  \bibinfo {author} {\bibfnamefont {V.}~\bibnamefont {Pruneri}}, \bibinfo
  {author} {\bibfnamefont {M.~W.}\ \bibnamefont {Mitchell}}, \bibinfo {author}
  {\bibfnamefont {M.}~\bibnamefont {Markham}}, \bibinfo {author} {\bibfnamefont
  {D.~J.}\ \bibnamefont {Twitchen}}, \bibinfo {author} {\bibfnamefont
  {D.}~\bibnamefont {Elkouss}}, \bibinfo {author} {\bibfnamefont
  {S.}~\bibnamefont {Wehner}}, \bibinfo {author} {\bibfnamefont {T.~H.}\
  \bibnamefont {Taminiau}},\ and\ \bibinfo {author} {\bibfnamefont
  {R.}~\bibnamefont {Hanson}},\ }\href
  {https://doi.org/10.4121/uuid:6e19e9b2-4a2d-40b5-8dd3-a660bf3c0a31} {\bibinfo
  {title} {Experimental loophole-free violation of a {{Bell}} inequality using
  entangled electron spins separated by 1.3 km}} (\bibinfo {year} {2015}),\
  \Eprint {https://arxiv.org/abs/1508.05949} {arXiv:1508.05949 [quant-ph]}
  \BibitemShut {NoStop}%
\bibitem [{\citenamefont {Giustina}\ \emph
  {et~al.}(2015{\natexlab{a}})\citenamefont {Giustina}, \citenamefont
  {Versteegh}, \citenamefont {Wengerowsky}, \citenamefont {Handsteiner},
  \citenamefont {Hochrainer}, \citenamefont {Phelan}, \citenamefont
  {Steinlechner}, \citenamefont {Kofler}, \citenamefont {Larsson},
  \citenamefont {Abell\'an}, \citenamefont {Amaya}, \citenamefont {Pruneri},
  \citenamefont {Mitchell}, \citenamefont {Beyer}, \citenamefont {Gerrits},
  \citenamefont {Lita}, \citenamefont {Shalm}, \citenamefont {Nam},
  \citenamefont {Scheidl}, \citenamefont {Ursin}, \citenamefont {Wittmann},\
  and\ \citenamefont {Zeilinger}}]{loopholefreephotons}%
  \BibitemOpen
  \bibfield  {author} {\bibinfo {author} {\bibfnamefont {M.}~\bibnamefont
  {Giustina}}, \bibinfo {author} {\bibfnamefont {M.~A.~M.}\ \bibnamefont
  {Versteegh}}, \bibinfo {author} {\bibfnamefont {S.}~\bibnamefont
  {Wengerowsky}}, \bibinfo {author} {\bibfnamefont {J.}~\bibnamefont
  {Handsteiner}}, \bibinfo {author} {\bibfnamefont {A.}~\bibnamefont
  {Hochrainer}}, \bibinfo {author} {\bibfnamefont {K.}~\bibnamefont {Phelan}},
  \bibinfo {author} {\bibfnamefont {F.}~\bibnamefont {Steinlechner}}, \bibinfo
  {author} {\bibfnamefont {J.}~\bibnamefont {Kofler}}, \bibinfo {author}
  {\bibfnamefont {J.-A.}\ \bibnamefont {Larsson}}, \bibinfo {author}
  {\bibfnamefont {C.}~\bibnamefont {Abell\'an}}, \bibinfo {author}
  {\bibfnamefont {W.}~\bibnamefont {Amaya}}, \bibinfo {author} {\bibfnamefont
  {V.}~\bibnamefont {Pruneri}}, \bibinfo {author} {\bibfnamefont {M.~W.}\
  \bibnamefont {Mitchell}}, \bibinfo {author} {\bibfnamefont {J.}~\bibnamefont
  {Beyer}}, \bibinfo {author} {\bibfnamefont {T.}~\bibnamefont {Gerrits}},
  \bibinfo {author} {\bibfnamefont {A.~E.}\ \bibnamefont {Lita}}, \bibinfo
  {author} {\bibfnamefont {L.~K.}\ \bibnamefont {Shalm}}, \bibinfo {author}
  {\bibfnamefont {S.~W.}\ \bibnamefont {Nam}}, \bibinfo {author} {\bibfnamefont
  {T.}~\bibnamefont {Scheidl}}, \bibinfo {author} {\bibfnamefont
  {R.}~\bibnamefont {Ursin}}, \bibinfo {author} {\bibfnamefont
  {B.}~\bibnamefont {Wittmann}},\ and\ \bibinfo {author} {\bibfnamefont
  {A.}~\bibnamefont {Zeilinger}},\ }\bibfield  {title} {\bibinfo {title}
  {Significant-loophole-free test of bell's theorem with entangled photons},\
  }\href {https://doi.org/10.1103/PhysRevLett.115.250401} {\bibfield  {journal}
  {\bibinfo  {journal} {Phys. Rev. Lett.}\ }\textbf {\bibinfo {volume} {115}},\
  \bibinfo {pages} {250401} (\bibinfo {year} {2015}{\natexlab{a}})}\BibitemShut
  {NoStop}%
\bibitem [{\citenamefont {Giustina}\ \emph
  {et~al.}(2015{\natexlab{b}})\citenamefont {Giustina}, \citenamefont
  {Versteegh}, \citenamefont {Wengerowsky}, \citenamefont {Handsteiner},
  \citenamefont {Hochrainer}, \citenamefont {Phelan}, \citenamefont
  {Steinlechner}, \citenamefont {Kofler}, \citenamefont {Larsson},
  \citenamefont {Abell\'an}, \citenamefont {Amaya}, \citenamefont {Pruneri},
  \citenamefont {Mitchell}, \citenamefont {Beyer}, \citenamefont {Gerrits},
  \citenamefont {Lita}, \citenamefont {Shalm}, \citenamefont {Nam},
  \citenamefont {Scheidl}, \citenamefont {Ursin}, \citenamefont {Wittmann},\
  and\ \citenamefont {Zeilinger}}]{significant}%
  \BibitemOpen
  \bibfield  {author} {\bibinfo {author} {\bibfnamefont {M.}~\bibnamefont
  {Giustina}}, \bibinfo {author} {\bibfnamefont {M.~A.~M.}\ \bibnamefont
  {Versteegh}}, \bibinfo {author} {\bibfnamefont {S.}~\bibnamefont
  {Wengerowsky}}, \bibinfo {author} {\bibfnamefont {J.}~\bibnamefont
  {Handsteiner}}, \bibinfo {author} {\bibfnamefont {A.}~\bibnamefont
  {Hochrainer}}, \bibinfo {author} {\bibfnamefont {K.}~\bibnamefont {Phelan}},
  \bibinfo {author} {\bibfnamefont {F.}~\bibnamefont {Steinlechner}}, \bibinfo
  {author} {\bibfnamefont {J.}~\bibnamefont {Kofler}}, \bibinfo {author}
  {\bibfnamefont {J.-A.}\ \bibnamefont {Larsson}}, \bibinfo {author}
  {\bibfnamefont {C.}~\bibnamefont {Abell\'an}}, \bibinfo {author}
  {\bibfnamefont {W.}~\bibnamefont {Amaya}}, \bibinfo {author} {\bibfnamefont
  {V.}~\bibnamefont {Pruneri}}, \bibinfo {author} {\bibfnamefont {M.~W.}\
  \bibnamefont {Mitchell}}, \bibinfo {author} {\bibfnamefont {J.}~\bibnamefont
  {Beyer}}, \bibinfo {author} {\bibfnamefont {T.}~\bibnamefont {Gerrits}},
  \bibinfo {author} {\bibfnamefont {A.~E.}\ \bibnamefont {Lita}}, \bibinfo
  {author} {\bibfnamefont {L.~K.}\ \bibnamefont {Shalm}}, \bibinfo {author}
  {\bibfnamefont {S.~W.}\ \bibnamefont {Nam}}, \bibinfo {author} {\bibfnamefont
  {T.}~\bibnamefont {Scheidl}}, \bibinfo {author} {\bibfnamefont
  {R.}~\bibnamefont {Ursin}}, \bibinfo {author} {\bibfnamefont
  {B.}~\bibnamefont {Wittmann}},\ and\ \bibinfo {author} {\bibfnamefont
  {A.}~\bibnamefont {Zeilinger}},\ }\bibfield  {title} {\bibinfo {title}
  {Significant-loophole-free test of bell's theorem with entangled photons},\
  }\href {https://doi.org/10.1103/PhysRevLett.115.250401} {\bibfield  {journal}
  {\bibinfo  {journal} {Phys. Rev. Lett.}\ }\textbf {\bibinfo {volume} {115}},\
  \bibinfo {pages} {250401} (\bibinfo {year} {2015}{\natexlab{b}})}\BibitemShut
  {NoStop}%
\bibitem [{\citenamefont {Shalm}\ \emph {et~al.}(2015)\citenamefont {Shalm},
  \citenamefont {Meyer-Scott}, \citenamefont {Christensen}, \citenamefont
  {Bierhorst}, \citenamefont {Wayne}, \citenamefont {Stevens}, \citenamefont
  {Gerrits}, \citenamefont {Glancy}, \citenamefont {Hamel}, \citenamefont
  {Allman}, \citenamefont {Coakley}, \citenamefont {Dyer}, \citenamefont
  {Hodge}, \citenamefont {Lita}, \citenamefont {Verma}, \citenamefont
  {Lambrocco}, \citenamefont {Tortorici}, \citenamefont {Migdall},
  \citenamefont {Zhang}, \citenamefont {Kumor}, \citenamefont {Farr},
  \citenamefont {Marsili}, \citenamefont {Shaw}, \citenamefont {Stern},
  \citenamefont {Abell\'an}, \citenamefont {Amaya}, \citenamefont {Pruneri},
  \citenamefont {Jennewein}, \citenamefont {Mitchell}, \citenamefont {Kwiat},
  \citenamefont {Bienfang}, \citenamefont {Mirin}, \citenamefont {Knill},\ and\
  \citenamefont {Nam}}]{strongtest}%
  \BibitemOpen
  \bibfield  {author} {\bibinfo {author} {\bibfnamefont {L.~K.}\ \bibnamefont
  {Shalm}}, \bibinfo {author} {\bibfnamefont {E.}~\bibnamefont {Meyer-Scott}},
  \bibinfo {author} {\bibfnamefont {B.~G.}\ \bibnamefont {Christensen}},
  \bibinfo {author} {\bibfnamefont {P.}~\bibnamefont {Bierhorst}}, \bibinfo
  {author} {\bibfnamefont {M.~A.}\ \bibnamefont {Wayne}}, \bibinfo {author}
  {\bibfnamefont {M.~J.}\ \bibnamefont {Stevens}}, \bibinfo {author}
  {\bibfnamefont {T.}~\bibnamefont {Gerrits}}, \bibinfo {author} {\bibfnamefont
  {S.}~\bibnamefont {Glancy}}, \bibinfo {author} {\bibfnamefont {D.~R.}\
  \bibnamefont {Hamel}}, \bibinfo {author} {\bibfnamefont {M.~S.}\ \bibnamefont
  {Allman}}, \bibinfo {author} {\bibfnamefont {K.~J.}\ \bibnamefont {Coakley}},
  \bibinfo {author} {\bibfnamefont {S.~D.}\ \bibnamefont {Dyer}}, \bibinfo
  {author} {\bibfnamefont {C.}~\bibnamefont {Hodge}}, \bibinfo {author}
  {\bibfnamefont {A.~E.}\ \bibnamefont {Lita}}, \bibinfo {author}
  {\bibfnamefont {V.~B.}\ \bibnamefont {Verma}}, \bibinfo {author}
  {\bibfnamefont {C.}~\bibnamefont {Lambrocco}}, \bibinfo {author}
  {\bibfnamefont {E.}~\bibnamefont {Tortorici}}, \bibinfo {author}
  {\bibfnamefont {A.~L.}\ \bibnamefont {Migdall}}, \bibinfo {author}
  {\bibfnamefont {Y.}~\bibnamefont {Zhang}}, \bibinfo {author} {\bibfnamefont
  {D.~R.}\ \bibnamefont {Kumor}}, \bibinfo {author} {\bibfnamefont {W.~H.}\
  \bibnamefont {Farr}}, \bibinfo {author} {\bibfnamefont {F.}~\bibnamefont
  {Marsili}}, \bibinfo {author} {\bibfnamefont {M.~D.}\ \bibnamefont {Shaw}},
  \bibinfo {author} {\bibfnamefont {J.~A.}\ \bibnamefont {Stern}}, \bibinfo
  {author} {\bibfnamefont {C.}~\bibnamefont {Abell\'an}}, \bibinfo {author}
  {\bibfnamefont {W.}~\bibnamefont {Amaya}}, \bibinfo {author} {\bibfnamefont
  {V.}~\bibnamefont {Pruneri}}, \bibinfo {author} {\bibfnamefont
  {T.}~\bibnamefont {Jennewein}}, \bibinfo {author} {\bibfnamefont {M.~W.}\
  \bibnamefont {Mitchell}}, \bibinfo {author} {\bibfnamefont {P.~G.}\
  \bibnamefont {Kwiat}}, \bibinfo {author} {\bibfnamefont {J.~C.}\ \bibnamefont
  {Bienfang}}, \bibinfo {author} {\bibfnamefont {R.~P.}\ \bibnamefont {Mirin}},
  \bibinfo {author} {\bibfnamefont {E.}~\bibnamefont {Knill}},\ and\ \bibinfo
  {author} {\bibfnamefont {S.~W.}\ \bibnamefont {Nam}},\ }\bibfield  {title}
  {\bibinfo {title} {Strong loophole-free test of local realism},\ }\href
  {https://doi.org/10.1103/PhysRevLett.115.250402} {\bibfield  {journal}
  {\bibinfo  {journal} {Phys. Rev. Lett.}\ }\textbf {\bibinfo {volume} {115}},\
  \bibinfo {pages} {250402} (\bibinfo {year} {2015})}\BibitemShut {NoStop}%
\bibitem [{\citenamefont {Brunner}\ \emph {et~al.}(2014)\citenamefont
  {Brunner}, \citenamefont {Cavalcanti}, \citenamefont {Pironio}, \citenamefont
  {Scarani},\ and\ \citenamefont {Wehner}}]{brunner}%
  \BibitemOpen
  \bibfield  {author} {\bibinfo {author} {\bibfnamefont {N.}~\bibnamefont
  {Brunner}}, \bibinfo {author} {\bibfnamefont {D.}~\bibnamefont {Cavalcanti}},
  \bibinfo {author} {\bibfnamefont {S.}~\bibnamefont {Pironio}}, \bibinfo
  {author} {\bibfnamefont {V.}~\bibnamefont {Scarani}},\ and\ \bibinfo {author}
  {\bibfnamefont {S.}~\bibnamefont {Wehner}},\ }\bibfield  {title} {\bibinfo
  {title} {Bell nonlocality},\ }\href
  {https://doi.org/10.1103/RevModPhys.86.419} {\bibfield  {journal} {\bibinfo
  {journal} {Rev. Mod. Phys.}\ }\textbf {\bibinfo {volume} {86}},\ \bibinfo
  {pages} {419} (\bibinfo {year} {2014})}\BibitemShut {NoStop}%
\bibitem [{\citenamefont {Acín}\ \emph {et~al.}(2006)\citenamefont {Acín},
  \citenamefont {Massar},\ and\ \citenamefont {Pironio}}]{acin2006}%
  \BibitemOpen
  \bibfield  {author} {\bibinfo {author} {\bibfnamefont {A.}~\bibnamefont
  {Acín}}, \bibinfo {author} {\bibfnamefont {S.}~\bibnamefont {Massar}},\ and\
  \bibinfo {author} {\bibfnamefont {S.}~\bibnamefont {Pironio}},\ }\bibfield
  {title} {\bibinfo {title} {Efficient quantum key distribution secure against
  no-signalling eavesdroppers},\ }\href
  {https://doi.org/10.1088/1367-2630/8/8/126} {\bibfield  {journal} {\bibinfo
  {journal} {New Journal of Physics}\ }\textbf {\bibinfo {volume} {8}},\
  \bibinfo {pages} {126} (\bibinfo {year} {2006})}\BibitemShut {NoStop}%
\bibitem [{\citenamefont {Ac\'{\i}n}\ \emph {et~al.}(2007)\citenamefont
  {Ac\'{\i}n}, \citenamefont {Brunner}, \citenamefont {Gisin}, \citenamefont
  {Massar}, \citenamefont {Pironio},\ and\ \citenamefont {Scarani}}]{acin2007}%
  \BibitemOpen
  \bibfield  {author} {\bibinfo {author} {\bibfnamefont {A.}~\bibnamefont
  {Ac\'{\i}n}}, \bibinfo {author} {\bibfnamefont {N.}~\bibnamefont {Brunner}},
  \bibinfo {author} {\bibfnamefont {N.}~\bibnamefont {Gisin}}, \bibinfo
  {author} {\bibfnamefont {S.}~\bibnamefont {Massar}}, \bibinfo {author}
  {\bibfnamefont {S.}~\bibnamefont {Pironio}},\ and\ \bibinfo {author}
  {\bibfnamefont {V.}~\bibnamefont {Scarani}},\ }\bibfield  {title} {\bibinfo
  {title} {Device-independent security of quantum cryptography against
  collective attacks},\ }\href {https://doi.org/10.1103/PhysRevLett.98.230501}
  {\bibfield  {journal} {\bibinfo  {journal} {Phys. Rev. Lett.}\ }\textbf
  {\bibinfo {volume} {98}},\ \bibinfo {pages} {230501} (\bibinfo {year}
  {2007})}\BibitemShut {NoStop}%
\bibitem [{\citenamefont {Pironio}\ \emph {et~al.}(2009)\citenamefont
  {Pironio}, \citenamefont {Acín}, \citenamefont {Brunner}, \citenamefont
  {Gisin}, \citenamefont {Massar},\ and\ \citenamefont {Scarani}}]{pironio}%
  \BibitemOpen
  \bibfield  {author} {\bibinfo {author} {\bibfnamefont {S.}~\bibnamefont
  {Pironio}}, \bibinfo {author} {\bibfnamefont {A.}~\bibnamefont {Acín}},
  \bibinfo {author} {\bibfnamefont {N.}~\bibnamefont {Brunner}}, \bibinfo
  {author} {\bibfnamefont {N.}~\bibnamefont {Gisin}}, \bibinfo {author}
  {\bibfnamefont {S.}~\bibnamefont {Massar}},\ and\ \bibinfo {author}
  {\bibfnamefont {V.}~\bibnamefont {Scarani}},\ }\bibfield  {title} {\bibinfo
  {title} {Device-independent quantum key distribution secure against
  collective attacks},\ }\href {https://doi.org/10.1088/1367-2630/11/4/045021}
  {\bibfield  {journal} {\bibinfo  {journal} {New Journal of Physics}\ }\textbf
  {\bibinfo {volume} {11}},\ \bibinfo {pages} {045021} (\bibinfo {year}
  {2009})}\BibitemShut {NoStop}%
\bibitem [{\citenamefont {Masanes}\ \emph {et~al.}(2011)\citenamefont
  {Masanes}, \citenamefont {Pironio},\ and\ \citenamefont
  {Ac{\'{\i}}n}}]{acin2011}%
  \BibitemOpen
  \bibfield  {author} {\bibinfo {author} {\bibfnamefont {L.}~\bibnamefont
  {Masanes}}, \bibinfo {author} {\bibfnamefont {S.}~\bibnamefont {Pironio}},\
  and\ \bibinfo {author} {\bibfnamefont {A.}~\bibnamefont {Ac{\'{\i}}n}},\
  }\bibfield  {title} {\bibinfo {title} {Secure device-independent quantum key
  distribution with causally independent measurement devices},\ }\bibfield
  {journal} {\bibinfo  {journal} {Nature Communications}\ }\textbf {\bibinfo
  {volume} {2}},\ \href {https://doi.org/10.1038/ncomms1244}
  {10.1038/ncomms1244} (\bibinfo {year} {2011})\BibitemShut {NoStop}%
\bibitem [{\citenamefont {Woodhead}\ \emph {et~al.}(2021)\citenamefont
  {Woodhead}, \citenamefont {Ac{\'{\i}}n},\ and\ \citenamefont
  {Pironio}}]{pironio2021}%
  \BibitemOpen
  \bibfield  {author} {\bibinfo {author} {\bibfnamefont {E.}~\bibnamefont
  {Woodhead}}, \bibinfo {author} {\bibfnamefont {A.}~\bibnamefont
  {Ac{\'{\i}}n}},\ and\ \bibinfo {author} {\bibfnamefont {S.}~\bibnamefont
  {Pironio}},\ }\bibfield  {title} {\bibinfo {title} {Device-independent
  quantum key distribution with asymmetric {CHSH} inequalities},\ }\href
  {https://doi.org/10.22331/q-2021-04-26-443} {\bibfield  {journal} {\bibinfo
  {journal} {Quantum}\ }\textbf {\bibinfo {volume} {5}},\ \bibinfo {pages}
  {443} (\bibinfo {year} {2021})}\BibitemShut {NoStop}%
\bibitem [{\citenamefont {Vazirani}\ and\ \citenamefont
  {Vidick}(2014)}]{vidick}%
  \BibitemOpen
  \bibfield  {author} {\bibinfo {author} {\bibfnamefont {U.}~\bibnamefont
  {Vazirani}}\ and\ \bibinfo {author} {\bibfnamefont {T.}~\bibnamefont
  {Vidick}},\ }\bibfield  {title} {\bibinfo {title} {Fully device-independent
  quantum key distribution},\ }\href
  {https://doi.org/10.1103/PhysRevLett.113.140501} {\bibfield  {journal}
  {\bibinfo  {journal} {Phys. Rev. Lett.}\ }\textbf {\bibinfo {volume} {113}},\
  \bibinfo {pages} {140501} (\bibinfo {year} {2014})}\BibitemShut {NoStop}%
\bibitem [{\citenamefont {Primaatmaja}\ \emph {et~al.}(2022)\citenamefont
  {Primaatmaja}, \citenamefont {Goh}, \citenamefont {Tan}, \citenamefont
  {Khoo}, \citenamefont {Ghorai},\ and\ \citenamefont {Lim}}]{diqkd_review1}%
  \BibitemOpen
  \bibfield  {author} {\bibinfo {author} {\bibfnamefont {I.~W.}\ \bibnamefont
  {Primaatmaja}}, \bibinfo {author} {\bibfnamefont {K.~T.}\ \bibnamefont
  {Goh}}, \bibinfo {author} {\bibfnamefont {E.~Y.~Z.}\ \bibnamefont {Tan}},
  \bibinfo {author} {\bibfnamefont {J.~T.~F.}\ \bibnamefont {Khoo}}, \bibinfo
  {author} {\bibfnamefont {S.}~\bibnamefont {Ghorai}},\ and\ \bibinfo {author}
  {\bibfnamefont {C.~C.~W.}\ \bibnamefont {Lim}},\ }\href@noop {} {\bibinfo
  {title} {Security of device-independent quantum key distribution protocols: a
  review}} (\bibinfo {year} {2022}),\ \Eprint
  {https://arxiv.org/abs/arXiv:2206.04960} {arXiv:2206.04960} \BibitemShut
  {NoStop}%
\bibitem [{\citenamefont {Zapatero}\ \emph {et~al.}(2022)\citenamefont
  {Zapatero}, \citenamefont {van Leent}, \citenamefont {Arnon-Friedman},
  \citenamefont {Liu}, \citenamefont {Zhang}, \citenamefont {Weinfurter},\ and\
  \citenamefont {Curty}}]{diqkd_review2}%
  \BibitemOpen
  \bibfield  {author} {\bibinfo {author} {\bibfnamefont {V.}~\bibnamefont
  {Zapatero}}, \bibinfo {author} {\bibfnamefont {T.}~\bibnamefont {van Leent}},
  \bibinfo {author} {\bibfnamefont {R.}~\bibnamefont {Arnon-Friedman}},
  \bibinfo {author} {\bibfnamefont {W.-Z.}\ \bibnamefont {Liu}}, \bibinfo
  {author} {\bibfnamefont {Q.}~\bibnamefont {Zhang}}, \bibinfo {author}
  {\bibfnamefont {H.}~\bibnamefont {Weinfurter}},\ and\ \bibinfo {author}
  {\bibfnamefont {M.}~\bibnamefont {Curty}},\ }\href@noop {} {\bibinfo {title}
  {Advances in device-independent quantum key distribution}} (\bibinfo {year}
  {2022}),\ \Eprint {https://arxiv.org/abs/arXiv:2208.12842} {arXiv:2208.12842}
  \BibitemShut {NoStop}%
\bibitem [{\citenamefont {Giovannetti}\ \emph {et~al.}(2011)\citenamefont
  {Giovannetti}, \citenamefont {Lloyd},\ and\ \citenamefont
  {Maccone}}]{metrology}%
  \BibitemOpen
  \bibfield  {author} {\bibinfo {author} {\bibfnamefont {V.}~\bibnamefont
  {Giovannetti}}, \bibinfo {author} {\bibfnamefont {S.}~\bibnamefont {Lloyd}},\
  and\ \bibinfo {author} {\bibfnamefont {L.}~\bibnamefont {Maccone}},\
  }\bibfield  {title} {\bibinfo {title} {Advances in quantum metrology},\
  }\href {https://doi.org/10.1038/nphoton.2011.35} {\bibfield  {journal}
  {\bibinfo  {journal} {Nature Photonics}\ }\textbf {\bibinfo {volume} {5}},\
  \bibinfo {pages} {222} (\bibinfo {year} {2011})}\BibitemShut {NoStop}%
\bibitem [{\citenamefont {Colombo}\ \emph {et~al.}(2022)\citenamefont
  {Colombo}, \citenamefont {Pedrozo-Pe{\~{n}}afiel}, \citenamefont
  {Adiyatullin}, \citenamefont {Li}, \citenamefont {Mendez}, \citenamefont
  {Shu},\ and\ \citenamefont {Vuleti{\'{c}}}}]{metrology2}%
  \BibitemOpen
  \bibfield  {author} {\bibinfo {author} {\bibfnamefont {S.}~\bibnamefont
  {Colombo}}, \bibinfo {author} {\bibfnamefont {E.}~\bibnamefont
  {Pedrozo-Pe{\~{n}}afiel}}, \bibinfo {author} {\bibfnamefont {A.~F.}\
  \bibnamefont {Adiyatullin}}, \bibinfo {author} {\bibfnamefont
  {Z.}~\bibnamefont {Li}}, \bibinfo {author} {\bibfnamefont {E.}~\bibnamefont
  {Mendez}}, \bibinfo {author} {\bibfnamefont {C.}~\bibnamefont {Shu}},\ and\
  \bibinfo {author} {\bibfnamefont {V.}~\bibnamefont {Vuleti{\'{c}}}},\
  }\bibfield  {title} {\bibinfo {title} {Time-reversal-based quantum metrology
  with many-body entangled states},\ }\href
  {https://doi.org/10.1038/s41567-022-01653-5} {\bibfield  {journal} {\bibinfo
  {journal} {Nature Physics}\ }\textbf {\bibinfo {volume} {18}},\ \bibinfo
  {pages} {925} (\bibinfo {year} {2022})}\BibitemShut {NoStop}%
\bibitem [{\citenamefont {Liu}\ \emph {et~al.}(2021{\natexlab{a}})\citenamefont
  {Liu}, \citenamefont {Zhang}, \citenamefont {Chen}, \citenamefont {Wang},\
  and\ \citenamefont {Yuan}}]{metrology3}%
  \BibitemOpen
  \bibfield  {author} {\bibinfo {author} {\bibfnamefont {J.}~\bibnamefont
  {Liu}}, \bibinfo {author} {\bibfnamefont {M.}~\bibnamefont {Zhang}}, \bibinfo
  {author} {\bibfnamefont {H.}~\bibnamefont {Chen}}, \bibinfo {author}
  {\bibfnamefont {L.}~\bibnamefont {Wang}},\ and\ \bibinfo {author}
  {\bibfnamefont {H.}~\bibnamefont {Yuan}},\ }\bibfield  {title} {\bibinfo
  {title} {Optimal scheme for quantum metrology},\ }\href
  {https://doi.org/10.1002/qute.202100080} {\bibfield  {journal} {\bibinfo
  {journal} {Advanced Quantum Technologies}\ }\textbf {\bibinfo {volume} {5}},\
  \bibinfo {pages} {2100080} (\bibinfo {year}
  {2021}{\natexlab{a}})}\BibitemShut {NoStop}%
\bibitem [{\citenamefont {Thekkadath}\ \emph {et~al.}(2020)\citenamefont
  {Thekkadath}, \citenamefont {Mycroft}, \citenamefont {Bell}, \citenamefont
  {Wade}, \citenamefont {Eckstein}, \citenamefont {Phillips}, \citenamefont
  {Patel}, \citenamefont {Buraczewski}, \citenamefont {Lita}, \citenamefont
  {Gerrits}, \citenamefont {Nam}, \citenamefont {Stobi{\'{n}}ska},
  \citenamefont {Lvovsky},\ and\ \citenamefont {Walmsley}}]{monikametrology}%
  \BibitemOpen
  \bibfield  {author} {\bibinfo {author} {\bibfnamefont {G.~S.}\ \bibnamefont
  {Thekkadath}}, \bibinfo {author} {\bibfnamefont {M.~E.}\ \bibnamefont
  {Mycroft}}, \bibinfo {author} {\bibfnamefont {B.~A.}\ \bibnamefont {Bell}},
  \bibinfo {author} {\bibfnamefont {C.~G.}\ \bibnamefont {Wade}}, \bibinfo
  {author} {\bibfnamefont {A.}~\bibnamefont {Eckstein}}, \bibinfo {author}
  {\bibfnamefont {D.~S.}\ \bibnamefont {Phillips}}, \bibinfo {author}
  {\bibfnamefont {R.~B.}\ \bibnamefont {Patel}}, \bibinfo {author}
  {\bibfnamefont {A.}~\bibnamefont {Buraczewski}}, \bibinfo {author}
  {\bibfnamefont {A.~E.}\ \bibnamefont {Lita}}, \bibinfo {author}
  {\bibfnamefont {T.}~\bibnamefont {Gerrits}}, \bibinfo {author} {\bibfnamefont
  {S.~W.}\ \bibnamefont {Nam}}, \bibinfo {author} {\bibfnamefont
  {M.}~\bibnamefont {Stobi{\'{n}}ska}}, \bibinfo {author} {\bibfnamefont
  {A.~I.}\ \bibnamefont {Lvovsky}},\ and\ \bibinfo {author} {\bibfnamefont
  {I.~A.}\ \bibnamefont {Walmsley}},\ }\bibfield  {title} {\bibinfo {title}
  {Quantum-enhanced interferometry with large heralded photon-number states},\
  }\bibfield  {journal} {\bibinfo  {journal} {npj Quantum Information}\
  }\textbf {\bibinfo {volume} {6}},\ \href
  {https://doi.org/10.1038/s41534-020-00320-y} {10.1038/s41534-020-00320-y}
  (\bibinfo {year} {2020})\BibitemShut {NoStop}%
\bibitem [{\citenamefont {Pironio}\ \emph {et~al.}(2010)\citenamefont
  {Pironio}, \citenamefont {Ac{\'{\i}}n}, \citenamefont {Massar}, \citenamefont
  {de~la Giroday}, \citenamefont {Matsukevich}, \citenamefont {Maunz},
  \citenamefont {Olmschenk}, \citenamefont {Hayes}, \citenamefont {Luo},
  \citenamefont {Manning},\ and\ \citenamefont {Monroe}}]{pironioRNG}%
  \BibitemOpen
  \bibfield  {author} {\bibinfo {author} {\bibfnamefont {S.}~\bibnamefont
  {Pironio}}, \bibinfo {author} {\bibfnamefont {A.}~\bibnamefont
  {Ac{\'{\i}}n}}, \bibinfo {author} {\bibfnamefont {S.}~\bibnamefont {Massar}},
  \bibinfo {author} {\bibfnamefont {A.~B.}\ \bibnamefont {de~la Giroday}},
  \bibinfo {author} {\bibfnamefont {D.~N.}\ \bibnamefont {Matsukevich}},
  \bibinfo {author} {\bibfnamefont {P.}~\bibnamefont {Maunz}}, \bibinfo
  {author} {\bibfnamefont {S.}~\bibnamefont {Olmschenk}}, \bibinfo {author}
  {\bibfnamefont {D.}~\bibnamefont {Hayes}}, \bibinfo {author} {\bibfnamefont
  {L.}~\bibnamefont {Luo}}, \bibinfo {author} {\bibfnamefont {T.~A.}\
  \bibnamefont {Manning}},\ and\ \bibinfo {author} {\bibfnamefont
  {C.}~\bibnamefont {Monroe}},\ }\bibfield  {title} {\bibinfo {title} {Random
  numbers certified by bell's theorem},\ }\href
  {https://doi.org/10.1038/nature09008} {\bibfield  {journal} {\bibinfo
  {journal} {Nature}\ }\textbf {\bibinfo {volume} {464}},\ \bibinfo {pages}
  {1021} (\bibinfo {year} {2010})}\BibitemShut {NoStop}%
\bibitem [{\citenamefont {Liu}\ \emph {et~al.}(2021{\natexlab{b}})\citenamefont
  {Liu}, \citenamefont {Li}, \citenamefont {Ragy}, \citenamefont {Zhao},
  \citenamefont {Bai}, \citenamefont {Liu}, \citenamefont {Brown},
  \citenamefont {Zhang}, \citenamefont {Colbeck}, \citenamefont {Fan},
  \citenamefont {Zhang},\ and\ \citenamefont {Pan}}]{RNG}%
  \BibitemOpen
  \bibfield  {author} {\bibinfo {author} {\bibfnamefont {W.-Z.}\ \bibnamefont
  {Liu}}, \bibinfo {author} {\bibfnamefont {M.-H.}\ \bibnamefont {Li}},
  \bibinfo {author} {\bibfnamefont {S.}~\bibnamefont {Ragy}}, \bibinfo {author}
  {\bibfnamefont {S.-R.}\ \bibnamefont {Zhao}}, \bibinfo {author}
  {\bibfnamefont {B.}~\bibnamefont {Bai}}, \bibinfo {author} {\bibfnamefont
  {Y.}~\bibnamefont {Liu}}, \bibinfo {author} {\bibfnamefont {P.~J.}\
  \bibnamefont {Brown}}, \bibinfo {author} {\bibfnamefont {J.}~\bibnamefont
  {Zhang}}, \bibinfo {author} {\bibfnamefont {R.}~\bibnamefont {Colbeck}},
  \bibinfo {author} {\bibfnamefont {J.}~\bibnamefont {Fan}}, \bibinfo {author}
  {\bibfnamefont {Q.}~\bibnamefont {Zhang}},\ and\ \bibinfo {author}
  {\bibfnamefont {J.-W.}\ \bibnamefont {Pan}},\ }\bibfield  {title} {\bibinfo
  {title} {Device-independent randomness expansion against quantum side
  information},\ }\href {https://doi.org/10.1038/s41567-020-01147-2} {\bibfield
   {journal} {\bibinfo  {journal} {Nature Physics}\ }\textbf {\bibinfo {volume}
  {17}},\ \bibinfo {pages} {448} (\bibinfo {year}
  {2021}{\natexlab{b}})}\BibitemShut {NoStop}%
\bibitem [{\citenamefont {Avesani}\ \emph {et~al.}(2021)\citenamefont
  {Avesani}, \citenamefont {Tebyanian}, \citenamefont {Villoresi},\ and\
  \citenamefont {Vallone}}]{RNG2}%
  \BibitemOpen
  \bibfield  {author} {\bibinfo {author} {\bibfnamefont {M.}~\bibnamefont
  {Avesani}}, \bibinfo {author} {\bibfnamefont {H.}~\bibnamefont {Tebyanian}},
  \bibinfo {author} {\bibfnamefont {P.}~\bibnamefont {Villoresi}},\ and\
  \bibinfo {author} {\bibfnamefont {G.}~\bibnamefont {Vallone}},\ }\bibfield
  {title} {\bibinfo {title} {Semi-device-independent heterodyne-based quantum
  random-number generator},\ }\href
  {https://doi.org/10.1103/PhysRevApplied.15.034034} {\bibfield  {journal}
  {\bibinfo  {journal} {Phys. Rev. Applied}\ }\textbf {\bibinfo {volume}
  {15}},\ \bibinfo {pages} {034034} (\bibinfo {year} {2021})}\BibitemShut
  {NoStop}%
\bibitem [{\citenamefont {Pearle}(1970)}]{fairsampling1}%
  \BibitemOpen
  \bibfield  {author} {\bibinfo {author} {\bibfnamefont {P.~M.}\ \bibnamefont
  {Pearle}},\ }\bibfield  {title} {\bibinfo {title} {Hidden-variable example
  based upon data rejection},\ }\href {https://doi.org/10.1103/PhysRevD.2.1418}
  {\bibfield  {journal} {\bibinfo  {journal} {Phys. Rev. D}\ }\textbf {\bibinfo
  {volume} {2}},\ \bibinfo {pages} {1418} (\bibinfo {year} {1970})}\BibitemShut
  {NoStop}%
\bibitem [{\citenamefont {Clauser}\ and\ \citenamefont
  {Horne}(1974)}]{fairsampling2}%
  \BibitemOpen
  \bibfield  {author} {\bibinfo {author} {\bibfnamefont {J.~F.}\ \bibnamefont
  {Clauser}}\ and\ \bibinfo {author} {\bibfnamefont {M.~A.}\ \bibnamefont
  {Horne}},\ }\bibfield  {title} {\bibinfo {title} {Experimental consequences
  of objective local theories},\ }\href
  {https://doi.org/10.1103/PhysRevD.10.526} {\bibfield  {journal} {\bibinfo
  {journal} {Phys. Rev. D}\ }\textbf {\bibinfo {volume} {10}},\ \bibinfo
  {pages} {526} (\bibinfo {year} {1974})}\BibitemShut {NoStop}%
\bibitem [{\citenamefont {Makarov}(2009)}]{detectorblinding1}%
  \BibitemOpen
  \bibfield  {author} {\bibinfo {author} {\bibfnamefont {V.}~\bibnamefont
  {Makarov}},\ }\bibfield  {title} {\bibinfo {title} {Controlling passively
  quenched single photon detectors by bright light},\ }\href
  {https://doi.org/10.1088/1367-2630/11/6/065003} {\bibfield  {journal}
  {\bibinfo  {journal} {New Journal of Physics}\ }\textbf {\bibinfo {volume}
  {11}},\ \bibinfo {pages} {065003} (\bibinfo {year} {2009})}\BibitemShut
  {NoStop}%
\bibitem [{\citenamefont {Lydersen}\ \emph {et~al.}(2010)\citenamefont
  {Lydersen}, \citenamefont {Wiechers}, \citenamefont {Wittmann}, \citenamefont
  {Elser}, \citenamefont {Skaar},\ and\ \citenamefont
  {Makarov}}]{detectorblinding2}%
  \BibitemOpen
  \bibfield  {author} {\bibinfo {author} {\bibfnamefont {L.}~\bibnamefont
  {Lydersen}}, \bibinfo {author} {\bibfnamefont {C.}~\bibnamefont {Wiechers}},
  \bibinfo {author} {\bibfnamefont {C.}~\bibnamefont {Wittmann}}, \bibinfo
  {author} {\bibfnamefont {D.}~\bibnamefont {Elser}}, \bibinfo {author}
  {\bibfnamefont {J.}~\bibnamefont {Skaar}},\ and\ \bibinfo {author}
  {\bibfnamefont {V.}~\bibnamefont {Makarov}},\ }\bibfield  {title} {\bibinfo
  {title} {Hacking commercial quantum cryptography systems by tailored bright
  illumination},\ }\href {https://doi.org/10.1038/nphoton.2010.214} {\bibfield
  {journal} {\bibinfo  {journal} {Nature Photonics}\ }\textbf {\bibinfo
  {volume} {4}},\ \bibinfo {pages} {686} (\bibinfo {year} {2010})}\BibitemShut
  {NoStop}%
\bibitem [{\citenamefont {Eberhard}(1993)}]{eberhard}%
  \BibitemOpen
  \bibfield  {author} {\bibinfo {author} {\bibfnamefont {P.~H.}\ \bibnamefont
  {Eberhard}},\ }\bibfield  {title} {\bibinfo {title} {Background level and
  counter efficiencies required for a loophole-free einstein-podolsky-rosen
  experiment},\ }\href {https://doi.org/10.1103/PhysRevA.47.R747} {\bibfield
  {journal} {\bibinfo  {journal} {Phys. Rev. A}\ }\textbf {\bibinfo {volume}
  {47}},\ \bibinfo {pages} {R747} (\bibinfo {year} {1993})}\BibitemShut
  {NoStop}%
\bibitem [{\citenamefont {Giustina}\ \emph {et~al.}(2013)\citenamefont
  {Giustina}, \citenamefont {Mech}, \citenamefont {Ramelow}, \citenamefont
  {Wittmann}, \citenamefont {Kofler}, \citenamefont {Beyer}, \citenamefont
  {Lita}, \citenamefont {Calkins}, \citenamefont {Gerrits}, \citenamefont
  {Nam}, \citenamefont {Ursin},\ and\ \citenamefont
  {Zeilinger}}]{closedetectionloophole}%
  \BibitemOpen
  \bibfield  {author} {\bibinfo {author} {\bibfnamefont {M.}~\bibnamefont
  {Giustina}}, \bibinfo {author} {\bibfnamefont {A.}~\bibnamefont {Mech}},
  \bibinfo {author} {\bibfnamefont {S.}~\bibnamefont {Ramelow}}, \bibinfo
  {author} {\bibfnamefont {B.}~\bibnamefont {Wittmann}}, \bibinfo {author}
  {\bibfnamefont {J.}~\bibnamefont {Kofler}}, \bibinfo {author} {\bibfnamefont
  {J.}~\bibnamefont {Beyer}}, \bibinfo {author} {\bibfnamefont
  {A.}~\bibnamefont {Lita}}, \bibinfo {author} {\bibfnamefont {B.}~\bibnamefont
  {Calkins}}, \bibinfo {author} {\bibfnamefont {T.}~\bibnamefont {Gerrits}},
  \bibinfo {author} {\bibfnamefont {S.~W.}\ \bibnamefont {Nam}}, \bibinfo
  {author} {\bibfnamefont {R.}~\bibnamefont {Ursin}},\ and\ \bibinfo {author}
  {\bibfnamefont {A.}~\bibnamefont {Zeilinger}},\ }\bibfield  {title} {\bibinfo
  {title} {Bell violation using entangled photons without the fair-sampling
  assumption},\ }\href {https://doi.org/10.1038/nature12012} {\bibfield
  {journal} {\bibinfo  {journal} {Nature}\ }\textbf {\bibinfo {volume} {497}},\
  \bibinfo {pages} {227} (\bibinfo {year} {2013})}\BibitemShut {NoStop}%
\bibitem [{\citenamefont {Zhang}\ \emph {et~al.}(2022)\citenamefont {Zhang},
  \citenamefont {van Leent}, \citenamefont {Redeker}, \citenamefont {Garthoff},
  \citenamefont {Schwonnek}, \citenamefont {Fertig}, \citenamefont {Eppelt},
  \citenamefont {Rosenfeld}, \citenamefont {Scarani}, \citenamefont {Lim},\
  and\ \citenamefont {Weinfurter}}]{diqkd}%
  \BibitemOpen
  \bibfield  {author} {\bibinfo {author} {\bibfnamefont {W.}~\bibnamefont
  {Zhang}}, \bibinfo {author} {\bibfnamefont {T.}~\bibnamefont {van Leent}},
  \bibinfo {author} {\bibfnamefont {K.}~\bibnamefont {Redeker}}, \bibinfo
  {author} {\bibfnamefont {R.}~\bibnamefont {Garthoff}}, \bibinfo {author}
  {\bibfnamefont {R.}~\bibnamefont {Schwonnek}}, \bibinfo {author}
  {\bibfnamefont {F.}~\bibnamefont {Fertig}}, \bibinfo {author} {\bibfnamefont
  {S.}~\bibnamefont {Eppelt}}, \bibinfo {author} {\bibfnamefont
  {W.}~\bibnamefont {Rosenfeld}}, \bibinfo {author} {\bibfnamefont
  {V.}~\bibnamefont {Scarani}}, \bibinfo {author} {\bibfnamefont {C.~C.-W.}\
  \bibnamefont {Lim}},\ and\ \bibinfo {author} {\bibfnamefont {H.}~\bibnamefont
  {Weinfurter}},\ }\bibfield  {title} {\bibinfo {title} {A device-independent
  quantum key distribution system for distant users},\ }\href
  {https://doi.org/10.1038/s41586-022-04891-y} {\bibfield  {journal} {\bibinfo
  {journal} {Nature}\ }\textbf {\bibinfo {volume} {607}},\ \bibinfo {pages}
  {687} (\bibinfo {year} {2022})}\BibitemShut {NoStop}%
\bibitem [{\citenamefont {Clauser}\ \emph {et~al.}(1969)\citenamefont
  {Clauser}, \citenamefont {Horne}, \citenamefont {Shimony},\ and\
  \citenamefont {Holt}}]{CHSH}%
  \BibitemOpen
  \bibfield  {author} {\bibinfo {author} {\bibfnamefont {J.~F.}\ \bibnamefont
  {Clauser}}, \bibinfo {author} {\bibfnamefont {M.~A.}\ \bibnamefont {Horne}},
  \bibinfo {author} {\bibfnamefont {A.}~\bibnamefont {Shimony}},\ and\ \bibinfo
  {author} {\bibfnamefont {R.~A.}\ \bibnamefont {Holt}},\ }\bibfield  {title}
  {\bibinfo {title} {Proposed {{Experiment}} to {{Test Local Hidden-Variable
  Theories}}},\ }\href {https://doi.org/10.1103/PhysRevLett.23.880} {\bibfield
  {journal} {\bibinfo  {journal} {Physical Review Letters}\ }\textbf {\bibinfo
  {volume} {23}},\ \bibinfo {pages} {880} (\bibinfo {year} {1969})}\BibitemShut
  {NoStop}%
\bibitem [{\citenamefont {Banaszek}\ and\ \citenamefont
  {W{\'o}dkiewicz}(1996)}]{bw0}%
  \BibitemOpen
  \bibfield  {author} {\bibinfo {author} {\bibfnamefont {K.}~\bibnamefont
  {Banaszek}}\ and\ \bibinfo {author} {\bibfnamefont {K.}~\bibnamefont
  {W{\'o}dkiewicz}},\ }\bibfield  {title} {\bibinfo {title} {Direct {{Probing}}
  of {{Quantum Phase Space}} by {{Photon Counting}}},\ }\href
  {https://doi.org/10.1103/PhysRevLett.76.4344} {\bibfield  {journal} {\bibinfo
   {journal} {Physical Review Letters}\ }\textbf {\bibinfo {volume} {76}},\
  \bibinfo {pages} {4344} (\bibinfo {year} {1996})}\BibitemShut {NoStop}%
\bibitem [{\citenamefont {Banaszek}\ and\ \citenamefont
  {W\'odkiewicz}(1998)}]{bw1}%
  \BibitemOpen
  \bibfield  {author} {\bibinfo {author} {\bibfnamefont {K.}~\bibnamefont
  {Banaszek}}\ and\ \bibinfo {author} {\bibfnamefont {K.}~\bibnamefont
  {W\'odkiewicz}},\ }\bibfield  {title} {\bibinfo {title} {Nonlocality of the
  einstein-podolsky-rosen state in the wigner representation},\ }\href
  {https://doi.org/10.1103/PhysRevA.58.4345} {\bibfield  {journal} {\bibinfo
  {journal} {Phys. Rev. A}\ }\textbf {\bibinfo {volume} {58}},\ \bibinfo
  {pages} {4345} (\bibinfo {year} {1998})}\BibitemShut {NoStop}%
\bibitem [{\citenamefont {Banaszek}\ and\ \citenamefont
  {W{\'o}dkiewicz}(1999)}]{bw2}%
  \BibitemOpen
  \bibfield  {author} {\bibinfo {author} {\bibfnamefont {K.}~\bibnamefont
  {Banaszek}}\ and\ \bibinfo {author} {\bibfnamefont {K.}~\bibnamefont
  {W{\'o}dkiewicz}},\ }\bibfield  {title} {\bibinfo {title} {Testing {{Quantum
  Nonlocality}} in {{Phase Space}}},\ }\href
  {https://doi.org/10.1103/PhysRevLett.82.2009} {\bibfield  {journal} {\bibinfo
   {journal} {Physical Review Letters}\ }\textbf {\bibinfo {volume} {82}},\
  \bibinfo {pages} {2009} (\bibinfo {year} {1999})}\BibitemShut {NoStop}%
\bibitem [{\citenamefont {Banaszek}\ \emph {et~al.}(2002)\citenamefont
  {Banaszek}, \citenamefont {Dragan}, \citenamefont {W\'odkiewicz},\ and\
  \citenamefont {Radzewicz}}]{bw3}%
  \BibitemOpen
  \bibfield  {author} {\bibinfo {author} {\bibfnamefont {K.}~\bibnamefont
  {Banaszek}}, \bibinfo {author} {\bibfnamefont {A.}~\bibnamefont {Dragan}},
  \bibinfo {author} {\bibfnamefont {K.}~\bibnamefont {W\'odkiewicz}},\ and\
  \bibinfo {author} {\bibfnamefont {C.}~\bibnamefont {Radzewicz}},\ }\bibfield
  {title} {\bibinfo {title} {Direct measurement of optical quasidistribution
  functions: Multimode theory and homodyne tests of bell's inequalities},\
  }\href {https://doi.org/10.1103/PhysRevA.66.043803} {\bibfield  {journal}
  {\bibinfo  {journal} {Phys. Rev. A}\ }\textbf {\bibinfo {volume} {66}},\
  \bibinfo {pages} {043803} (\bibinfo {year} {2002})}\BibitemShut {NoStop}%
\bibitem [{\citenamefont {Jeong}\ \emph {et~al.}(2003)\citenamefont {Jeong},
  \citenamefont {Son}, \citenamefont {Kim}, \citenamefont {Ahn},\ and\
  \citenamefont {Brukner}}]{bwtmsv}%
  \BibitemOpen
  \bibfield  {author} {\bibinfo {author} {\bibfnamefont {H.}~\bibnamefont
  {Jeong}}, \bibinfo {author} {\bibfnamefont {W.}~\bibnamefont {Son}}, \bibinfo
  {author} {\bibfnamefont {M.~S.}\ \bibnamefont {Kim}}, \bibinfo {author}
  {\bibfnamefont {D.}~\bibnamefont {Ahn}},\ and\ \bibinfo {author}
  {\bibfnamefont {i.~c.~v.}\ \bibnamefont {Brukner}},\ }\bibfield  {title}
  {\bibinfo {title} {Quantum nonlocality test for continuous-variable states
  with dichotomic observables},\ }\href
  {https://doi.org/10.1103/PhysRevA.67.012106} {\bibfield  {journal} {\bibinfo
  {journal} {Phys. Rev. A}\ }\textbf {\bibinfo {volume} {67}},\ \bibinfo
  {pages} {012106} (\bibinfo {year} {2003})}\BibitemShut {NoStop}%
\bibitem [{\citenamefont {Lee}\ and\ \citenamefont {Jeong}(2011)}]{bwcglmp}%
  \BibitemOpen
  \bibfield  {author} {\bibinfo {author} {\bibfnamefont {S.-W.}\ \bibnamefont
  {Lee}}\ and\ \bibinfo {author} {\bibfnamefont {H.}~\bibnamefont {Jeong}},\
  }\bibfield  {title} {\bibinfo {title} {High-dimensional bell test for a
  continuous-variable state in phase space and its robustness to detection
  inefficiency},\ }\href {https://doi.org/10.1103/PhysRevA.83.022103}
  {\bibfield  {journal} {\bibinfo  {journal} {Phys. Rev. A}\ }\textbf {\bibinfo
  {volume} {83}},\ \bibinfo {pages} {022103} (\bibinfo {year}
  {2011})}\BibitemShut {NoStop}%
\bibitem [{\citenamefont {Dastidar}\ and\ \citenamefont
  {Sarbicki}(2022)}]{detectingentanglement}%
  \BibitemOpen
  \bibfield  {author} {\bibinfo {author} {\bibfnamefont {M.~G.}\ \bibnamefont
  {Dastidar}}\ and\ \bibinfo {author} {\bibfnamefont {G.}~\bibnamefont
  {Sarbicki}},\ }\bibfield  {title} {\bibinfo {title} {Detecting entanglement
  between modes of light},\ }\href
  {https://doi.org/10.1103/PhysRevA.105.062459} {\bibfield  {journal} {\bibinfo
   {journal} {Phys. Rev. A}\ }\textbf {\bibinfo {volume} {105}},\ \bibinfo
  {pages} {062459} (\bibinfo {year} {2022})}\BibitemShut {NoStop}%
\bibitem [{\citenamefont {Brask}\ and\ \citenamefont {Chaves}(2012)}]{brask}%
  \BibitemOpen
  \bibfield  {author} {\bibinfo {author} {\bibfnamefont {J.~B.}\ \bibnamefont
  {Brask}}\ and\ \bibinfo {author} {\bibfnamefont {R.}~\bibnamefont {Chaves}},\
  }\bibfield  {title} {\bibinfo {title} {Robust nonlocality tests with
  displacement-based measurements},\ }\href
  {https://doi.org/10.1103/PhysRevA.86.010103} {\bibfield  {journal} {\bibinfo
  {journal} {Phys. Rev. A}\ }\textbf {\bibinfo {volume} {86}},\ \bibinfo
  {pages} {010103} (\bibinfo {year} {2012})}\BibitemShut {NoStop}%
\bibitem [{\citenamefont {Mycroft}\ \emph {et~al.}(2022)\citenamefont
  {Mycroft}, \citenamefont {McDermott}, \citenamefont {Buraczewski},\ and\
  \citenamefont {Stobi{\'n}ska}}]{mycroftProposalDistributionMultiphoton2022}%
  \BibitemOpen
  \bibfield  {author} {\bibinfo {author} {\bibfnamefont {M.~E.}\ \bibnamefont
  {Mycroft}}, \bibinfo {author} {\bibfnamefont {T.}~\bibnamefont {McDermott}},
  \bibinfo {author} {\bibfnamefont {A.}~\bibnamefont {Buraczewski}},\ and\
  \bibinfo {author} {\bibfnamefont {M.}~\bibnamefont {Stobi{\'n}ska}},\ }\href
  {https://doi.org/10.48550/arXiv.1812.10935} {\bibinfo {title} {Proposal for
  distribution of multi-photon entanglement with {{TF-QKD}} rate-distance
  scaling}} (\bibinfo {year} {2022}),\ \Eprint
  {https://arxiv.org/abs/1812.10935} {arXiv:1812.10935 [quant-ph]} \BibitemShut
  {NoStop}%
\bibitem [{\citenamefont {Ketterer}(2016)}]{ketterer}%
  \BibitemOpen
  \bibfield  {author} {\bibinfo {author} {\bibfnamefont {A.}~\bibnamefont
  {Ketterer}},\ }\emph {\bibinfo {title} {Modular Variables in Quantum
  Information}},\ \href@noop {} {Ph.D. thesis},\ \bibinfo  {school}
  {Universit\'e Paris 7, Sorbonne Paris Cit\'e} (\bibinfo {year}
  {2016})\BibitemShut {NoStop}%
\bibitem [{\citenamefont {Collins}\ \emph {et~al.}(2002)\citenamefont
  {Collins}, \citenamefont {Gisin}, \citenamefont {Linden}, \citenamefont
  {Massar},\ and\ \citenamefont {Popescu}}]{cglmp}%
  \BibitemOpen
  \bibfield  {author} {\bibinfo {author} {\bibfnamefont {D.}~\bibnamefont
  {Collins}}, \bibinfo {author} {\bibfnamefont {N.}~\bibnamefont {Gisin}},
  \bibinfo {author} {\bibfnamefont {N.}~\bibnamefont {Linden}}, \bibinfo
  {author} {\bibfnamefont {S.}~\bibnamefont {Massar}},\ and\ \bibinfo {author}
  {\bibfnamefont {S.}~\bibnamefont {Popescu}},\ }\bibfield  {title} {\bibinfo
  {title} {Bell inequalities for arbitrarily high-dimensional systems},\ }\href
  {https://doi.org/10.1103/PhysRevLett.88.040404} {\bibfield  {journal}
  {\bibinfo  {journal} {Phys. Rev. Lett.}\ }\textbf {\bibinfo {volume} {88}},\
  \bibinfo {pages} {040404} (\bibinfo {year} {2002})}\BibitemShut {NoStop}%
\bibitem [{\citenamefont {Banaszek}\ \emph {et~al.}(1999)\citenamefont
  {Banaszek}, \citenamefont {Radzewicz}, \citenamefont {W\'odkiewicz},\ and\
  \citenamefont {Krasi\ifmmode~\acute{n}\else \'{n}\fi{}ski}}]{wignerbypnr}%
  \BibitemOpen
  \bibfield  {author} {\bibinfo {author} {\bibfnamefont {K.}~\bibnamefont
  {Banaszek}}, \bibinfo {author} {\bibfnamefont {C.}~\bibnamefont {Radzewicz}},
  \bibinfo {author} {\bibfnamefont {K.}~\bibnamefont {W\'odkiewicz}},\ and\
  \bibinfo {author} {\bibfnamefont {J.~S.}\ \bibnamefont
  {Krasi\ifmmode~\acute{n}\else \'{n}\fi{}ski}},\ }\bibfield  {title} {\bibinfo
  {title} {Direct measurement of the wigner function by photon counting},\
  }\href {https://doi.org/10.1103/PhysRevA.60.674} {\bibfield  {journal}
  {\bibinfo  {journal} {Phys. Rev. A}\ }\textbf {\bibinfo {volume} {60}},\
  \bibinfo {pages} {674} (\bibinfo {year} {1999})}\BibitemShut {NoStop}%
\bibitem [{\citenamefont {Nehra}\ \emph {et~al.}(2019)\citenamefont {Nehra},
  \citenamefont {Win}, \citenamefont {Eaton}, \citenamefont {Shahrokhshahi},
  \citenamefont {Sridhar}, \citenamefont {Gerrits}, \citenamefont {Lita},
  \citenamefont {Nam},\ and\ \citenamefont {Pfister}}]{wignerbypnr2}%
  \BibitemOpen
  \bibfield  {author} {\bibinfo {author} {\bibfnamefont {R.}~\bibnamefont
  {Nehra}}, \bibinfo {author} {\bibfnamefont {A.}~\bibnamefont {Win}}, \bibinfo
  {author} {\bibfnamefont {M.}~\bibnamefont {Eaton}}, \bibinfo {author}
  {\bibfnamefont {R.}~\bibnamefont {Shahrokhshahi}}, \bibinfo {author}
  {\bibfnamefont {N.}~\bibnamefont {Sridhar}}, \bibinfo {author} {\bibfnamefont
  {T.}~\bibnamefont {Gerrits}}, \bibinfo {author} {\bibfnamefont
  {A.}~\bibnamefont {Lita}}, \bibinfo {author} {\bibfnamefont {S.~W.}\
  \bibnamefont {Nam}},\ and\ \bibinfo {author} {\bibfnamefont {O.}~\bibnamefont
  {Pfister}},\ }\bibfield  {title} {\bibinfo {title} {State-independent quantum
  state tomography by photon-number-resolving measurements},\ }\href
  {https://doi.org/10.1364/OPTICA.6.001356} {\bibfield  {journal} {\bibinfo
  {journal} {Optica}\ }\textbf {\bibinfo {volume} {6}},\ \bibinfo {pages}
  {1356} (\bibinfo {year} {2019})}\BibitemShut {NoStop}%
\bibitem [{\citenamefont {Devetak}\ and\ \citenamefont
  {Winter}(2005)}]{devetakwinter}%
  \BibitemOpen
  \bibfield  {author} {\bibinfo {author} {\bibfnamefont {I.}~\bibnamefont
  {Devetak}}\ and\ \bibinfo {author} {\bibfnamefont {A.}~\bibnamefont
  {Winter}},\ }\bibfield  {title} {\bibinfo {title} {Distillation of secret key
  and entanglement from quantum states},\ }\href
  {https://doi.org/10.1098/rspa.2004.1372} {\bibfield  {journal} {\bibinfo
  {journal} {Proceedings of the Royal Society A: Mathematical, Physical and
  Engineering Sciences}\ }\textbf {\bibinfo {volume} {461}},\ \bibinfo {pages}
  {207} (\bibinfo {year} {2005})}\BibitemShut {NoStop}%
\end{thebibliography}%


%


\end{document}


\title{Supplementary Information:\\ Eberhard limit for photon-counting Bell tests and its utility in quantum key distribution}

\author{Thomas McDermott}
\email{tommcdee78@gmail.com}
\thanks{Corresponding author}
\author{Morteza Moradi}
\author{Antoni Mikos-Nuszkiewicz}
\author{Magdalena Stobi\'nska}

\date{\today}

\maketitle

\tableofcontents

\newpage

\section*{Supplementary Note 1: Photon-counting CGLMP inequality}

\begin{figure}
    \centering
    \includegraphics[width=.65\linewidth]{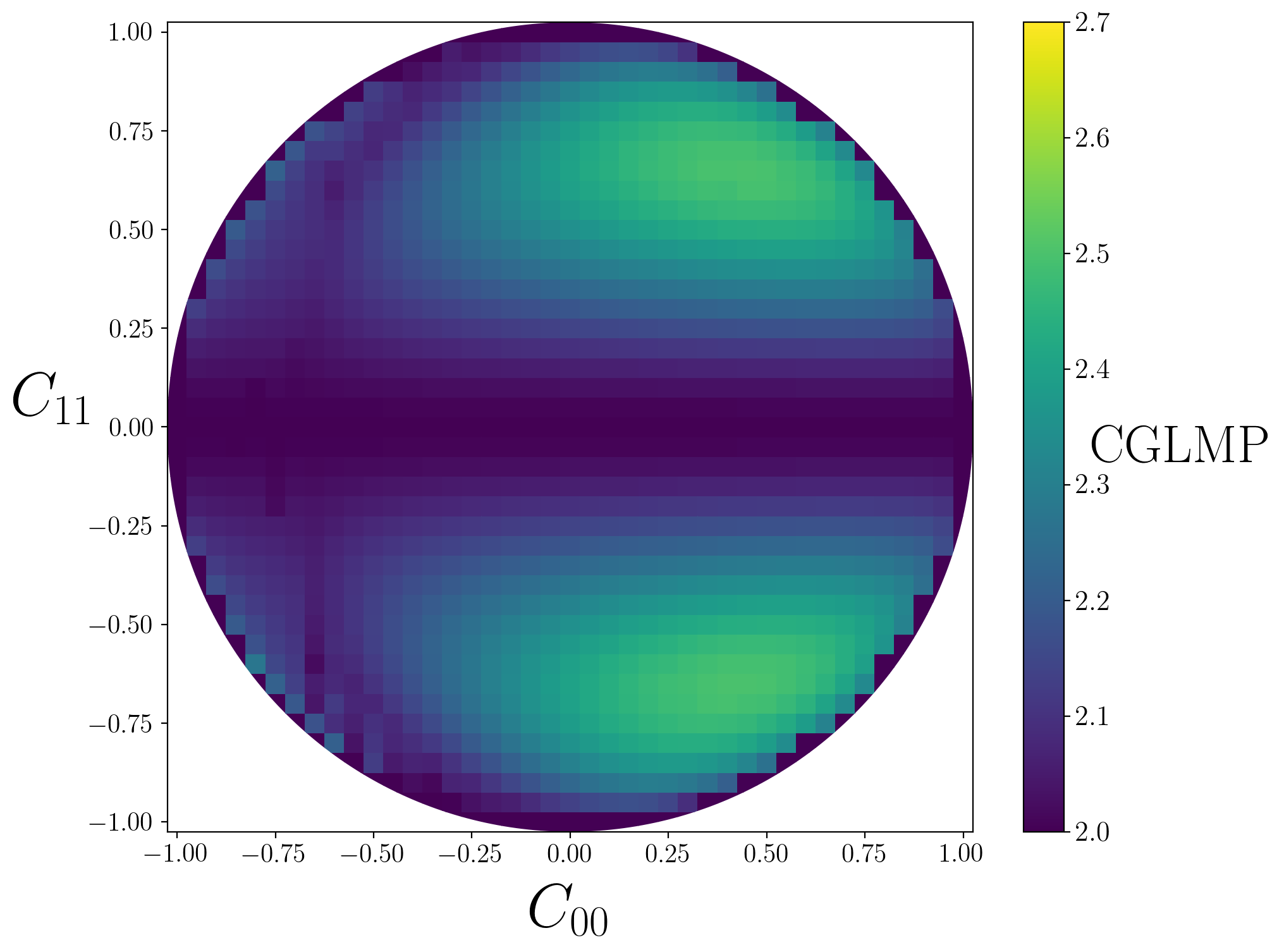}
    \caption{Results of CGLMP photon-counting Bell test for a state $C_{00}\ket{00} + C_{11}\ket{11} + C_{22}\ket{22}$, with $C_{22}$ fixed by normalization. The maximum value $\approx 2.50$ is not as high as the maximum value, $2.69$, for the zero/non-zero CHSH test.}
    \label{figS1}
\end{figure}

Consider a Bell test where Alice and Bob use two measurement settings each, $\delta_{\alpha 1}, \delta_{\alpha 2}$ and $\delta_{\beta 1}, \delta_{\beta 2}$, but an arbitrary number of outcomes $\Delta$. The facet-defining Bell inequalities in this scenario are the CGLMP inequalities which may be expressed as the following combination of probabilities
\begin{align}
	I = &+[p(i = j|\delta_{\alpha 1}, \delta_{\beta 1}) - p(i = j-1 |\delta_{\alpha 1}, \delta_{\beta 1})]\nonumber\\
	&-[p(i = j|\delta_{\alpha 2}, \delta_{\beta 1}) - p(i = j-1 |\delta_{\alpha 2}, \delta_{\beta 1})]\nonumber\\
	&+[p(i = j|\delta_{\alpha 1}, \delta_{\beta 2}) - p(i = j-1 |\delta_{\alpha 1}, \delta_{\beta 2})]\nonumber\\
	&+[p(i = j|\delta_{\alpha 1}, \delta_{\beta 2}) - p(j = i-1 |\delta_{\alpha 1}, \delta_{\beta 2})],
\end{align}
$I \leq 2$, where $i, j$ are Alice and Bob's measurement outcomes listed from $0$ to $\Delta - 1$, and the equalities are understood to be taken modulo $\Delta$. Notice here that the first three square bracketed terms have the same form, while the final one is the same except for $i \leftrightarrow j$. For $\Delta = 2$ this reduces to CHSH, so let us take the simplest non-trivial case $\Delta = 3$. In our scenario, Alice and Bob's measurements are photon number measurements, where the detection of two or more photons are grouped together into the same outcome. The terms we must calculate are then
\begin{align}
	p(i = j) &= p(i=0, j=0) + p(i=1, j=1) + p(i\geq2, j\geq2),\\
	p(i=j-1) &= p(i=0, j=1) + p(i=1, j\geq 2) + p(i\geq2, j=0). 
\end{align}
The terms $p(i=1, j\geq 2)$, $p(i\geq2, j=0)$ and $p(i\geq2, j\geq2)$ are inconvenient as they require summing probabilities of an infinite number of photon numbers. Luckily they can be simplified to finite sums
\begin{align}
	p(i=1, j\geq 2) &= p(i=1) - p(i=1, j=0) - p(i=1, j=1),\\
	p(i\geq 2, j = 0) &= p(j=0) - p(i=0, j=0) - p(i=1, j=0),\\
	p(i\geq 2, j\geq 2) &= 1 - p(i=0) - p(i=1) - p(j=0) - p(j=1)\nonumber\\
	&\hspace{1em} + p(i=0, j=0) + p(i=1, j=1) + p(i=1, j=0) + p(i=0, j=1).
\end{align}
Thus we can define
\begin{align}
	G(\delta_{\alpha}, \delta_{\beta}) &= p(i = j|\delta_{\alpha}, \delta_{\beta}) - p(i = j-1 |\delta_{\alpha}, \delta_{\beta})\\
	&= 3\Big[p(i=0,j=0|\delta_\alpha,\delta_\beta) \nonumber\\
	&\hspace{1em}+ p(i=1,j=0|\delta_\alpha,\delta_\beta)\nonumber\\ &\hspace{1em}+ p(i=1,j=1|\delta_\alpha,\delta_\beta)\Big]\nonumber\\ &\hspace{1em}-2p(i=1|\delta_\alpha) -2p(j=0|\delta_\beta)\nonumber\\&\hspace{1em}- p(i=0|\delta_{\alpha}) - p(j = 1|\delta_{\beta}) + 1,\nonumber
\end{align}
and the photon-counting CGLMP inequality becomes
\begin{equation}
	I = G(\delta_{\alpha 1}, \delta_{\beta 1}) - G(\delta_{\alpha 2}, \delta_{\beta 1}) + G(\delta_{\alpha 2}, \delta_{\beta 2}) + G(\delta_{\beta 2}, \delta_{\alpha 1}).
\end{equation}
The optimal CGLMP violation for a two-photon entangled state $C_{00}\ket{00} + C_{11}\ket{11} + C_{22}\ket{22}$ is shown in Fig. \ref{figS1}. The results are very similar to the even/odd CHSH test, but with slightly higher amounts of Bell violation. However, the maximum value obtained $\approx 2.50$ is not as that obtained, $2.69$, for the zero/non-zero CHSH test

\newpage
\section*{Supplementary Note 2: Key rate equations}

Here for convenience we outline a simple proof\cite{pironio2021} for the lower bound of the Devetak-Winter rate\cite{devetakwinter} used in the main paper:
\begin{equation}\label{KeyRateExpression}
    K_{DW}=H(B_1|E)-H(B_1|A_0) \geq 1 - \phi\big(\sqrt{S^2/4-1}\big)-H(B_1|A_0),
\end{equation}
where $\phi(x):=h(\frac{1+x}{2})$ and and $h$ is the binary entropy. Using the fact that any generalized measurement is projective in a larger Hilbert space, we can assume that Alice's and Bob's measurements are von Neumann measurements. Furthermore, since Alice and Bob both perform binary measurements, using the Jordan lemma, we can reduce the problem to qubit systems.
We know the measurements are projective. Thus the CHSH inequality cannot be violated if any of the measurements are  $\pm\mathbb{I}$ and they must all be linear combinations of the three orthogonal Pauli operators. We can define measurements by the angles between them on the Bloch sphere which is basis-independent. So we assume the local bases are as follows
\begin{equation}\label{angleA}
    A_1+A_2 = 2\cos\left(\frac{\theta_A}{2}\right) A \hspace{2mm},\hspace{2mm} A_1-A_2 = 2\sin\left(\frac{\theta_A}{2}\right) \Bar{A},
\end{equation}
\begin{equation}\label{angleB}
        B_1 = B \hspace{2mm},\hspace{2mm} B_2 = \cos(\theta_B) B + \sin(\theta_B) \Bar{B},
\end{equation}

\vspace{2mm}

\noindent where $\{A,\Bar{A}\}$ and $\{B,\Bar{B}\}$ are any pairs of orthogonal Pauli operators and  $\theta_A$ and $\theta_B$ are unknown angles. Inserting Eqs. (\ref{angleA}),(\ref{angleB}) into the CHSH expression we obtain
\begin{align}
        S & = \braket{(A_1+A_2) \otimes B_1}+\braket{(A_1-A_2) \otimes B_2}\nonumber\\
        & = 2\cos\left(\frac{\theta_A}{2}\right) \braket{A\otimes B} + 2\sin\left(\frac{\theta_A}{2}\right) \cos(\theta_B) \braket{\Bar{A}\otimes B} + 2\sin\left(\frac{\theta_A}{2}\right) \sin(\theta_B) \braket{\Bar{A}\otimes \Bar{B}}.
 \end{align}
Using the Cauchy-Schwarz inequality and the fact that
\begin{equation}
\cos\left(\frac{\theta_A}{2}\right)^2+\Big[\sin\left(\frac{\theta_A}{2}\right) \cos(\theta_B)\Big]^2+ \Big[\sin\left(\frac{\theta_A}{2}\right) \sin(\theta_B)\Big]^2=1.
\end{equation}
the following upper bound for $S$ can be obtained
\begin{equation}\label{Cauchy}
    S\leq 2\sqrt{\braket{A\otimes B}^2 +\braket{\Bar{A}\otimes B}^2 +\braket{\Bar{A}\otimes \Bar{B}}^2}.
\end{equation}
Combining Eq. (\ref{Cauchy}) with lemma \ref{CorrelationBound} we have
\begin{equation}\label{Correlation,CHSH}
    S \leq 2\sqrt{1+\braket{\Bar{A}\otimes \Bar{B}}^2} \hspace{2mm}\Rightarrow\hspace{2mm} |\braket{\Bar{A}\otimes \Bar{B}}|\geq \sqrt{S^2/4-1}.
\end{equation}
Finally, putting the lower bound given by Eq. (\ref{Correlation,CHSH}) in lemma \ref{EntropyBound} and using the monotonicity of function $\phi$, we have
\begin{equation}
H(B_1|E)\geq 1-\phi\big{(}\sqrt{S^2/4-1}\big{)}
\end{equation}
which gives us the lower bound on the key rate in Eq. (\ref{KeyRateExpression}).

\begin{lemma}\label{CorrelationBound}
\textbf{Correlation Bound.}
\begin{equation}
\braket{A\otimes B}^2 +\braket{\Bar{A}\otimes B}^2 \leq 1,
\end{equation}
where $\{A,\Bar{A}\}$ are two orthogonal Pauli observable (i.e. $\pm 1$ valued hermitian operator) on Alice's system and $B$ is any given Pauli observable on Bob's system.
\end{lemma}

\vspace{1mm}

\begin{proof} Since $\Vec{\sigma}:=(A , \Bar{A})$ consists of two orthogonal Pauli operators, for normalized vectors $\bold{a}=(a,a')^T , \bold{b}=(b,b')^T \in \mathbb{R}^2$, the combinations $\bold{a}.\Vec{\sigma}:= aA+a'\Bar{A}$ and $\bold{b}.\Vec{\sigma}:= bA+b'\Bar{A}$ are also $\pm 1$ valued hermitian operators, which results in $\braket{\bold{a}.\Vec{\sigma} \otimes \bold{b}.\Vec{\sigma}}\leq1$. Defining correlation matrix
\begin{equation}\label{aCb}
     C:=
     \begin{bmatrix}
      \braket{A\otimes B} &\braket{A\otimes \Bar{B}}\\
      \braket{\Bar{A}\otimes B} &\braket{\Bar{A}\otimes \Bar{B}}
     \end{bmatrix}
\Rightarrow \bold{a}^T C \bold{b} = \braket{\bold{a}.\Vec{\sigma} \otimes \bold{b}.\Vec{\sigma}}\leq1.
\end{equation}
We can remove the normalization condition by replacing $\bold{a}:={\bold{x}/\|\bold{x}\|}$ and $\bold{b}:=\bold{y}/\|\bold{y}\|$ in Eq. (\ref{aCb}), which result in $\bold{x}^T C \bold{y}\leq \|\bold{x}\| \|\bold{y}\|$. Choosing $\bold{x}=C\bold{y}$ we have
\begin{equation}\label{NormCompare}
    \|\bold{x}\|^2=\bold{x}^T\bold{x}=\bold{x}^T C \bold{y} \leq \|\bold{x}\|.\|\bold{y}\| \Rightarrow \|\bold{x}\|\leq \|\bold{y}\|,
\end{equation}
and also
\begin{equation}\label{CCNorm}
    \bold{y}^T C^T C \bold{y}=\bold{x}^T C \bold{y}\leq \|\bold{x}\|.\|\bold{y}\|\leq \|\bold{y}\|^2.
\end{equation}
where we used Eq. (\ref{NormCompare}) in the last inequality. Finally, the proof will be completed by choosing $\bold{y}^T=(y,y')=(1,0)$ and using Eq. (\ref{CCNorm})
\begin{equation}
\bold{y}^T C^T C \bold{y}=(C^TC)_{11}=\braket{A\otimes B}^2 +\braket{\Bar{A}\otimes B}^2\leq \|\bold{y}\|^2=1.
\end{equation}
\end{proof}

\vspace{1mm}

\begin{lemma}\label{EntropyBound} \textbf{Entropy Bound.} 
\begin{equation}
H(B|E)\geq 1-\phi\big{(}|\braket{A\otimes \Bar{B}}|\big{)},
\end{equation}
where $\{B,\Bar{B}\}$ are two orthogonal Pauli observable (i.e. $\pm 1$ valued hermitian operator) on Bob's system and $A$ is any given Pauli observable on Alice's system.
\end{lemma}

\vspace{1mm}

\begin{proof}
We know that by extension of Eve's system and purifying the initial state the conditional entropy cannot increase. So we can assume a pure state $\ket{\Psi}_{ABE}$ is shared  between  Alice, Bob, and Eve:
\begin{equation}\label{PsiABE}
    \ket{\Psi}_{ABE}=\ket{0}_B\otimes\ket{\psi_0}_{AE}+\ket{1}_B\otimes\ket{\psi_1}_{AE},
\end{equation}
where $\ket{0}_B$ and $\ket{1}_B$ are the eigenstates of $B$. To calculate $H(B|E)$ we need to find the correlation between Bob and Eve after Bob measures B, which can be determined using the classical-quantum state
$\tau_{BE}=[0]_B\otimes\psi^E_0+[1]_B\otimes\psi^E_1$
where $\psi^E_x=Tr_A \ket{\psi_x}\bra{\psi_x}$:
\begin{equation}\label{EntropyBE}
    H(B|E)_\tau=S(\tau_{BE})-S(\tau_E)=S(\psi^E_0)+S(\psi^E_1)-S(\psi^E_0+\psi^E_1).
\end{equation}

\noindent Now lets compute the conditional entropy $H(B|CE)$ on the state
\begin{equation}\label{TauBCE}
\tau'_{BCE} = \frac{1}{2}[0]_B \otimes ([0]_C\otimes \psi^E_0 + [1]_C\otimes \psi^E_1)  + \frac{1}{2}[1]_B \otimes ([0]_C\otimes \psi^E_1 + [1]_C\otimes \psi^E_0),
\end{equation}
which is equal to
\begin{equation}\label{EntropyBCE}
    H(B|CE)_{\tau'}=S(\tau'_{BCE})-S(\tau'_{CE})= 2S\big{(}\frac{\psi^E_0}{2}\big{)}+2S\big{(}\frac{\psi^E_1}{2}\big{)}-2S\big{(}\frac{\psi^E_0+\psi^E_1}{2}\big{)}.    
\end{equation}

\noindent Using $2S(X/2)=S(X)+Tr(X)$, one can conclude from Eqs. (\ref{EntropyBE}) and (\ref{EntropyBCE}) that
\begin{equation}\label{ChangeTau}
    H(B|E)_\tau=H(B|CE)_{\tau'}.
\end{equation}

\noindent Since $H(B|CE)\geq H(B|ACDE)$ for any extended state $\tau'_{ABCDE}$ such that $Tr_{AD}[\tau'_{ABCDE}]=\tau'_{BCE}$, we can consider the following extension of $\tau'_{BCE}$:
\begin{equation}\label{TauABCDE}
\tau'_{ABCDE}=\frac{1}{2}[0]_B \otimes \ket{X_0}\bra{X_0}  + \frac{1}{2}[1]_B \otimes \ket{X_1}\bra{X_1},
\end{equation}

\noindent where we replaced the state of subsystem $CE$ in Eq. (\ref{TauBCE}) with purifications
\begin{equation}
\ket{X_0}=\ket{\psi_0}_{AE}\otimes\ket{00}_{CD}+(A\otimes \mathbb{I}_E)\ket{\psi_1}_{AE}\otimes\ket{11}_{CD},
\end{equation}
\begin{equation}
\ket{X_1}=(A\otimes \mathbb{I}_E) \ket{\psi_1}_{AE}\otimes\ket{00}_{CD} +\ket{\psi_0}_{AE}\otimes\ket{11}_{CD},
\end{equation}
where A is any $\pm 1$ valued hermitian operator on Alice’s system. Thus
\begin{equation}\label{EntropyBACDE}
    \begin{split}
         H(B|ACDE)_{\tau' }
         & = S(\tau'_{ABCDE})-S(\tau'_{ACDE})\\
         & = S\big{(}\frac{\ket{X_0}\bra{X_0}}{2}\big{)}+S\big{(}\frac{\ket{X_1}\bra{X_1}}{2}\big{)}-S\big{(}\frac{\ket{X_0}\bra{X_0}+\ket{X_1}\bra{X_1}}{2}\big{)}\\
         & = \frac{1}{2}+\frac{1}{2}-\phi\big{(}|\braket{X_0|X_1}|\big{)}.\\
    \end{split}
\end{equation}

\noindent We use Eq. (\ref{PsiABE}) to calculate $\braket{X_0|X_1}$:
\begin{equation}\label{Fidelity}
    \begin{split}
        \braket{X_0|X_1} & = \bra{\psi_0} A\otimes \mathbb{I}_E \ket{\psi_1} + \bra{\psi_1} A\otimes \mathbb{I}_E \ket{\psi_0} = \bra{\Psi} A\otimes \Bar{B} \otimes\mathbb{I}_E \ket{\Psi}_{ABE}\\
        & = Tr_{AB}Tr_E\big{[}(A\otimes\Bar{B}\otimes\mathbb{I}_E)\ket{\Psi}\bra{\Psi}\big{]} =\braket{A\otimes\Bar{B}}.
    \end{split}
\end{equation}

\noindent Finally, the proof will be completed by combining Eqs. (\ref{ChangeTau}), (\ref{EntropyBACDE}), (\ref{Fidelity}):
\begin{equation}
H(B|E)_\tau=H(B|CE)_{\tau'}\geq H(B|ACDE))_{\tau'}= 1-\phi\big{(}|\braket{A\otimes\Bar{B}}|\big{)}.
\end{equation}
\end{proof}

\vspace{3mm}

\subsection*{Comparison of Key rates}
There exists another lower bound for the key rate proposed in \cite{pironio}:
\begin{equation}\label{SymmetrizedKeyRate}
    K_{DW}\geq 1-h(Q)-\phi\big{(}\sqrt{S^2/4-1}\big{)},
\end{equation}
where h is the binary entropy and $Q$ is the quantum bit error rate (QBER) defined as $P(a\neq b|A_0,B_1)$.
Here we analytically show that the lower bound Eq. (\ref{KeyRateExpression}) always yields us a higher key rate than Eq. (\ref{SymmetrizedKeyRate}). To this end, it is enough to show that $h(Q)\geq H(B_1|A_0)$. Defining $P_{ij}=P(a=i,b=j|A_0,B_1)$ and $\delta=P_{00}P_{10}-P_{01}P_{11}$, we can calculate the difference of $h(Q)$ and $H(B_1|A_0)$:
\begin{equation}
    \begin{split}
        & \Delta := h(Q) - H(B_1|A_0) = h(P_{01}+P_{10})+h(P_{00}+P_{01})-\sum_{ij} P_{ij}\log(P_{ij})\\
        & = -P_{00}\log_2 \big{(}1-\frac{\delta_{}}{P_{00}}\big{)} -P_{10}\log_2 \big{(}1-\frac{\delta_{}}{P_{10}}\big{)} -P_{01}\log_2 \big{(}1+\frac{\delta_{}}{P_{01}}\big{)} -P_{11}\log_2 \big{(}1+\frac{\delta_{}}{P_{11}}\big{)}.
    \end{split}
\end{equation}
Using the Taylor series expansion $\log_e{(1+x)}=\sum (-1)^{n+1}{x^n}/{n}$, we have

\begin{equation}\label{Delta}
\begin{split}
        \Delta
        & =(\log_e^2)^{-1}. \sum_{n=2}^{\infty} \frac{\delta^n}{n}\big{[}\big{(}\frac{1}{P_{00}}\big{)}^{n-1}+\big{(}\frac{1}{P_{10}}\big{)}^{n-1}-\big{(}\frac{-1}{P_{01}}\big{)}^{n-1}-\big{(}\frac{-1}{P_{11}}\big{)}^{n-1}\big{]}\\
        & =(\log_e^2)^{-1}.\Big{[}\sum_{x,y\in\{0,1\}} P_{xy}\Delta_{xy}\Big{]} \hspace{3mm};\hspace{3mm} \Delta_{xy}:=\sum_{n=2}^\infty\frac{1}{n}\big{(}(-1)^{y}\frac{\delta}{P_{xy}}\big{)}^n.
\end{split}
\end{equation}
Without loss of generality, we can assume $\delta\geq 0$. Otherwise we can redefine the positive variable $\delta'=-\delta=P_{11}P_{01}-P_{10}P_{00}$ and Eq. (\ref{Delta}) will change to
\begin{equation}\Delta=(\log_e^2)^{-1}. \sum_{n=2}^{\infty} \frac{\delta'^n}{n}\big{[}\big{(}\frac{1}{P_{11}}\big{)}^{n-1}+\big{(}\frac{1}{P_{01}}\big{)}^{n-1}-\big{(}\frac{-1}{P_{10}}\big{)}^{n-1}-\big{(}\frac{-1}{P_{00}}\big{)}^{n-1}\big{]},
\end{equation}
which upon relabeling probabilities indices, is the same as Eq. (\ref{Delta}).

Since $\delta$ is positive, $\Delta_{x0}=\sum_{n=2}^\infty\big{(}\frac{\delta}{P_{x1}}\big{)}^n/n\geq 0$. Thus, to prove the positivity of $\Delta$, it is enough to show that $\Delta_{x1}=\sum_{n=2}^\infty\big{(}\frac{-\delta}{P_{x1}}\big{)}^n/n\geq 0$. To this end, we consider the value of probability $P_{x1}$ in comparison with $\delta$. There are 3 general cases:

\vspace{3mm}

\noindent\textbf{Case I)} $P_{x1} < {\delta}/{2}$:
\begin{equation}
    \begin{split}
        & \Rightarrow \forall k\in\mathbb{N}: P_{x1} < \frac{2k-1}{2k} \delta
        \Rightarrow \frac{1}{2k-1}\big{(}\frac{-\delta}{P_{x1}}\big{)}^{2k-1}+\frac{1}{2k}\big{(}\frac{-\delta}{P_{x1}}\big{)}^{2k} > 0,\\
        & \Rightarrow \Delta_{x1} = \frac{1}{2}\big{(}\frac{\delta}{P_{x1}}\big{)}^{2} + \sum_{k=2}^{\infty}\Big{[}\frac{1}{2k-1}\big{(}\frac{-\delta}{P_{x1}}\big{)}^{2k-1}+\frac{1}{2k}\big{(}\frac{-\delta}{P_{x1}}\big{)}^{2k}\Big{]} > 0,
    \end{split}
\end{equation}

\noindent\textbf{Case II)} $P_{x1}\geq \delta$:
\begin{equation}
    \begin{split}
        & \Rightarrow \forall k\in\mathbb{N}: P_{x1} \geq \frac{2k}{2k+1} \delta
        \Rightarrow \frac{1}{2k}\big{(}\frac{-\delta}{P_{x1}}\big{)}^{2k}+\frac{1}{2k+1}\big{(}\frac{-\delta}{P_{x1}}\big{)}^{2k+1}\geq 0, \\
        & \Rightarrow \Delta_{x1} = \sum_{k=1}^{\infty} \Big{[} \frac{1}{2k}\big{(}\frac{-\delta}{P_{x1}}\big{)}^{2k}+\frac{1}{2k+1}\big{(}\frac{-\delta}{P_{x1}}\big{)}^{2k+1} \Big{]}\geq 0 ,
    \end{split}
\end{equation}

\noindent\textbf{Case III)} $\delta/{2} \leq P_{x1} < {\delta}$:
\begin{equation}
    \Rightarrow \hspace{2mm} \exists k'\in\mathbb{N}: \frac{2k'-1}{2k'} \delta \leq  P_{x1} < \frac{2k'+1}{2k'+2} \delta,
\end{equation}
which gives us Lower-Bound(LB) and Upper-Bound(UB) of $P_{x1}$, from which we can infer some inequalities:

\begin{equation}\label{LB}
         LB \Rightarrow
          \forall k< k':  P_{x1} > \frac{2k}{2k+1} \delta
          \Rightarrow \frac{1}{2k}\big{(}\frac{-\delta}{P_{x1}}\big{)}^{2k}+\frac{1}{2k+1}\big{(}\frac{-\delta}{P_{x1}}\big{)}^{2k+1} > 0\\
\end{equation}

\begin{equation}\label{UB}
    UB \Rightarrow
         \forall k\geq k':  P_{x1} < \frac{2k+1}{2k+2} \delta
        \Rightarrow \frac{1}{2k+1}\big{(}\frac{-\delta}{P_{x1}}\big{)}^{2k+1}+\frac{1}{2k+2}\big{(}\frac{-\delta}{P_{x1}}\big{)}^{2k+2} > 0\\
\end{equation}

\noindent Writing $\Delta_{x1}$ in terms of Eqs. (\ref{LB}) and (\ref{UB}), we have
\begin{equation}
    \begin{split}
         \Delta_{x1} & = \sum_{k=1}^{k'-1} \Big{[} \frac{1}{2k}\big{(}\frac{-\delta}{P_{x1}}\big{)}^{2k}+\frac{1}{2k+1}\big{(}\frac{-\delta}{P_{x1}}\big{)}^{2k+1} \Big{]} + \frac{1}{2k'}\big{(}\frac{\delta}{P_{x1}}\big{)}^{2k'}\\
        & + \sum_{k=k'}^{\infty}\Big{[}\frac{1}{2k+1}\big{(}\frac{-\delta}{P_{x1}}\big{)}^{2k+1}+\frac{1}{2k+2}\big{(}\frac{-\delta}{P_{x1}}\big{)}^{2k+2}\Big{]} > 0. 
    \end{split}
\end{equation}\hfill\(\Box\)

We proved that for all probabilities $P_{xy}$, the value of $\Delta_{xy}$ is not negative. Therefore
\begin{equation}
\Delta=h(Q)-H(B_1|A_0)=(\log_e^2)^{-1}.\big{[}\sum P_{xy}\Delta_{xy}\big{]}\geq 0,
\end{equation}
or equivalently, the lower bound Eq. (\ref{KeyRateExpression}) is higher than Eq. (\ref{SymmetrizedKeyRate}) and equality happens only if $\delta=P_{00}P_{10}-P_{01}P_{11}$.

\bibliography{supplementary}